\documentclass[11pt,a4paper]{article}
\pdfoutput=1
\usepackage{jcappub}
\usepackage{bm}
\usepackage{amsmath}
\usepackage{color}

\newcommand{\Lya}{Ly$\alpha$~}
\newcommand{\hMpc}{h^{-1}~\mathrm{Mpc}}
\newcommand{\angstrom}{\text{\normalfont\AA}}

\title{The triply-ionized carbon forest from eBOSS: cosmological correlations with quasars in SDSS-IV DR14}

\author[a]{Michael~Blomqvist,}
\author[a]{Matthew~M.~Pieri,}
\author[b]{H{\'e}lion~du~Mas~des~Bourboux,}
\author[c]{Nicol{\'a}s~G.~Busca,}
\author[d]{An\v{z}e~Slosar,}
\author[b]{Julian~E.~Bautista,}
\author[e]{Jonathan~Brinkmann,}
\author[b]{Joel~R.~Brownstein,}
\author[b]{Kyle~Dawson,}
\author[c]{Victoria~de~Sainte~Agathe,}
\author[f]{Julien~Guy,}
\author[g]{Will~J.~Percival,}
\author[a]{Ignasi~P{\'e}rez-R{\`a}fols,}
\author[h]{James~Rich,}
\author[i,j]{Donald~P.~Schneider}

\affiliation[a]{Aix Marseille Univ, CNRS, LAM, Laboratoire d'Astrophysique de Marseille,\\ Marseille, France}
\affiliation[b]{Department of Physics and Astronomy, University of Utah,\\ 115 S. 1400 E., Salt Lake City, UT 84112, U.S.A.}
\affiliation[c]{LPNHE, CNRS/IN2P3, Universit{\'e} Pierre et Marie Curie Paris 6,\\ Universit{\'e} Denis Diderot Paris 7, 4 place Jussieu, 75252 Paris CEDEX, France}
\affiliation[d]{Brookhaven National Laboratory,\\ Upton NY 11973, U.S.A.}
\affiliation[e]{Apache Point Observatory,\\ P.O. Box 59, Sunspot, NM 88349, U.S.A.}
\affiliation[f]{Lawrence Berkeley National Laboratory,\\ 1 Cyclotron Road, Berkeley, CA 94720, U.S.A.}
\affiliation[g]{Institute of Cosmology \& Gravitation, University of Portsmouth,\\ Dennis Sciama Building, Portsmouth, PO1 3FX, UK}
\affiliation[h]{IRFU, CEA, Universit{\'e} Paris-Saclay,\\ F-91191 Gif-sur-Yvette, France}
\affiliation[i]{Department of Astronomy and Astrophysics, The Pennsylvania State University,\\
University Park, PA 16802, U.S.A.}
\affiliation[j]{Institute for Gravitation and the Cosmos, The Pennsylvania State University,\\
University Park, PA 16802, U.S.A.}

\emailAdd{michael.blomqvist@lam.fr, matthew.pieri@lam.fr, h.du.mas.des.bourboux@utah.edu, nbusca@lpnhe.in2p3.fr, anze@bnl.gov, bautista@astro.utah.edu, jb@apo.nmsu.edu, joelbrownstein@astro.utah.edu, kdawson@astro.utah.edu, victoria.de.sainte.agathe@lpnhe.in2p3.fr, jguy@lbl.gov, will.percival@port.ac.uk, ignasi.perez@lam.fr, james.rich@cea.fr, dps7@psu.edu}

\abstract{We present measurements of the cross-correlation of the triply-ionized carbon (CIV) forest with quasars using Sloan Digital Sky Survey Data Release 14. The study exploits a large sample of new quasars from the first two years of observations by the Extended Baryon Oscillation Spectroscopic Survey (eBOSS). The CIV forest is a weaker tracer of large-scale structure than the Ly$\alpha$ forest, but benefits from being accessible at redshifts $z<2$ where the quasar number density from eBOSS is high. Our data sample consists of 287,651 CIV forest quasars in the redshift range $1.4<z<3.5$ and 387,315 tracer quasars with $1.2<z<3.5$. We measure large-scale correlations from CIV absorption occuring in three distinct quasar rest-frame wavelength bands of the spectra referred to as the CIV forest, the SiIV forest and the Ly$\alpha$ forest. From the combined fit to the quasar-CIV cross-correlations for the CIV forest and the SiIV forest, the CIV redshift-space distortion parameter is $\beta_{\rm CIV}=0.27_{\ -0.14\ -0.26}^{\ +0.16\ +0.34}$ and its combination with the CIV linear transmission bias parameter is $b_{\rm CIV}(1+\beta_{\rm CIV})=-0.0183_{\ -0.0014\ -0.0029}^{\ +0.0013\ +0.0025}$ ($1\sigma$ and $2\sigma$ statistical errors) at the mean redshift $z=2.00$. Splitting the sample at $z=2.2$ to constrain the bias evolution with redshift yields the power-law exponent $\gamma=0.60\pm0.63$, indicating a significantly weaker redshift-evolution than for the Ly$\alpha$ forest linear transmission bias. Additionally, we demonstrate that CIV absorption has the potential to be used as a probe of baryon acoustic oscillations (BAO). While the current data set is insufficient for a detection of the BAO peak feature, the final quasar samples for redshifts $1.4<z<2.2$ from eBOSS and the Dark Energy Spectroscopic Instrument (DESI) are expected to provide measurements of the isotropic BAO scale to $\sim7\%$ and $\sim3\%$ precision, respectively, at $z\simeq1.6$.}


\begin{document}
\maketitle


\section{Introduction}

Intergalactic gas traces the distribution of matter on large scales and is measured in absorption along the line of sight to background quasars \cite{1965ApJ...142.1633G,1971ApJ...164L..73L, 1998ApJ...495...44C,1998ASPC..148...21W}. The use of these absorption features as a tracer of large-scale structure has become an important cosmological probe with the advent of spectroscopic surveys of high-redshift quasars. This measurement is predominantly achieved using the Lyman-$\alpha$ (Ly$\alpha$) resonant transition of neutral hydrogen. The Baryon Oscillation Spectroscopic Survey \cite{2013AJ....145...10D} (BOSS) of the Sloan Digital Sky Survey III \cite{2011AJ....142...72E,2006AJ....131.2332G} (SDSS-III) has provided an unprecedented number of spectra and (crucially) high number density of quasars with $z>2.1$, enabling the three-dimensional measurement of large-scale structure in the \Lya forest for the first time.

Studies using BOSS quasars provided the first detection of baryon acoustic oscillations (BAO) in the three-dimensional Ly$\alpha$ forest auto-correlation function at redshift $z\simeq2.3$ \cite{2013A&A...552A..96B,2013JCAP...04..026S,2013JCAP...03..024K}, extending the use of BAO as a probe of the cosmic expansion history \cite{2013PhR...530...87W} into the epoch of decelerated expansion when the Universe was strongly matter dominated \cite{2003ApJ...585...34M,2007PhRvD..76f3009M}. The measurements of the BAO scale were subsequently improved \cite{2015A&A...574A..59D} and complemented with a BAO detection in the cross-correlation of Ly$\alpha$ forest absorption with quasars at a similar redshift \cite{2014JCAP...05..027F}. Combining the measurements of the \Lya forest auto-correlation and cross-correlation with quasars from the completed SDSS-III high-redshift quasar sample available in SDSS Data Release 12 \cite{2015ApJS..219...12A} (DR12) constrained the position of the BAO peak in the radial and transverse directions to a precision of 2.5\% and 3.3\%, respectively \cite{2017A&A...603A..12B,2017A&A...608A.130D}.

The Extended Baryon Oscillation Spectroscopic Survey \cite{2016AJ....151...44D} (eBOSS) of the Sloan Digital Sky Survey IV \cite{2017AJ....154...28B} (SDSS-IV) continues the effort to observe quasars over a wide redshift range, with the goal of gathering a sample of 500,000 spectroscopically-confirmed quasars in the range $0.9<z<2.2$ for the purpose of studying quasar clustering, as well as providing 60,000 new quasars and reobservations of 60,000 BOSS quasars at $z>2.1$ for \Lya forest analyses. The first two years of eBOSS observations were released in SDSS Data Release 14 \cite{2017arXiv170709322A} (DR14). Using quasars in the range $0.8<z<2.2$ from this data set, a first measurement of BAO in quasar clustering was reported in \cite{2017arXiv170506373A}.

Other elements, traditionally dubbed `metals', also produce line-of-sight absorption in optical quasar spectroscopy \cite{1987ApJ...315L...5M, 2004MNRAS.347..985P,2006ApJ...638...45P}. Whether they have escaped their parent dark matter halos and joined the intergalactic medium or remain bound to the source galaxy in the circumgalactic medium, they can generate metal `forests', albeit weaker and more sparse than the \Lya forests. While  these metal forests are observed in BOSS data \cite{2014MNRAS.441.1718P}, they have not previously been used as large-scale structure tracers since this sample lacked the number density on the sky and/or the survey lacked sufficient area to produce meaningful results. With the advent of the eBOSS survey (plus its pilot survey SEQUELS; see section~\ref{sec:data}) sufficient number densities are being reached on quasars with $z<2.2$. Since the \Lya transition drops out of the optical window for quasars with $z\lesssim 2$, the metal lines detected in eBOSS provide a valuable opportunity to perform large-scale quasar absorption cosmology at these lower redshifts. Triply-ionized carbon (CIV) is the most promising candidate \cite{2014MNRAS.445L.104P}: it is the strongest tracer of structure amongst metals due to the high relative abundance of carbon, a high proportion of which is in the triply-ionized state due to the extragalactic UV background and the physical conditions found in the intergalactic medium \cite{1997ApJ...481..601R}. It also has convenient strong doublet resonance transitions with rest-frame wavelengths $1548.20~\angstrom$ and $1550.78~\angstrom$ on the red side of the \Lya forest. This property allows CIV to be observable down to $z\approx1.4$ in optical spectroscopy and be the dominant absorber for rest-frame wavelengths $>\lambda_{\rm Ly\alpha}=1215.67~\angstrom$. The SiIV forest (at wavelengths $<1394~\angstrom$ in the quasar rest-frame) does add significant contaminating absorption, but it remains subdominant to CIV \cite{2014MNRAS.445L.104P}.

There have been previous studies sensitive to the large-scale clustering of CIV. Some studies have measured the large-scale association between CIV and galaxies \cite{2003ApJ...584...45A, 2006ApJ...638...45P,2014MNRAS.445..794T} in order to probe the properties of galactic outflows. Other studies have attempted to probe the large-scale association between quasars and CIV absorption on megaparsec scales \cite{2009MNRAS.392.1539T,2014ApJ...796..140P}. Finally, the large-scale clustering of a catalogue of CIV absorbers has been cross-correlated with quasar positions \cite{2013ApJ...768...38V}, which are a stronger subset of the data we study (given the fact that lines in the catalogue must be individually statistically significant and confirmed with other metal lines).

We present results of cross-correlating the CIV forest with quasars positions, taking advantage of the high bias of quasars as a boost to the CIV tracer signal. This measurement is performed two years into the six-year eBOSS survey. We perform a first measurement of BAO using CIV forest absorption and explore the quantities driving the measured large-scale distribution of CIV. We further describe projections of expected BAO precision for the end of eBOSS and the Dark Energy Spectroscopic Instrument (DESI) experiment \cite{2016arXiv161100036D} (which will obtain spectra of 1.7 million $z<2.1$ quasars).

This paper is organized as follows: Section~\ref{sec:data} describes the DR14 data. Our method for measuring the flux-transmission field is explained in section~\ref{sec:transmission}. Section~\ref{sec:1Dcorr} presents a measurement of the one-dimensional correlation function. Section~\ref{sec:crosscorrelation} describes our method for measuring the cross-correlation function. The model that is used to fit the cross-correlation function is introduced in section~\ref{sec:model}. Our results are presented in section~\ref{sec:results} and discussed in section~\ref{sec:discussion}. The paper concludes with a summary in section~\ref{sec:summary}.

\section{Data samples and reduction}
\label{sec:data}

\begin{figure}
   \begin{center}
   \includegraphics[width=5.0in]{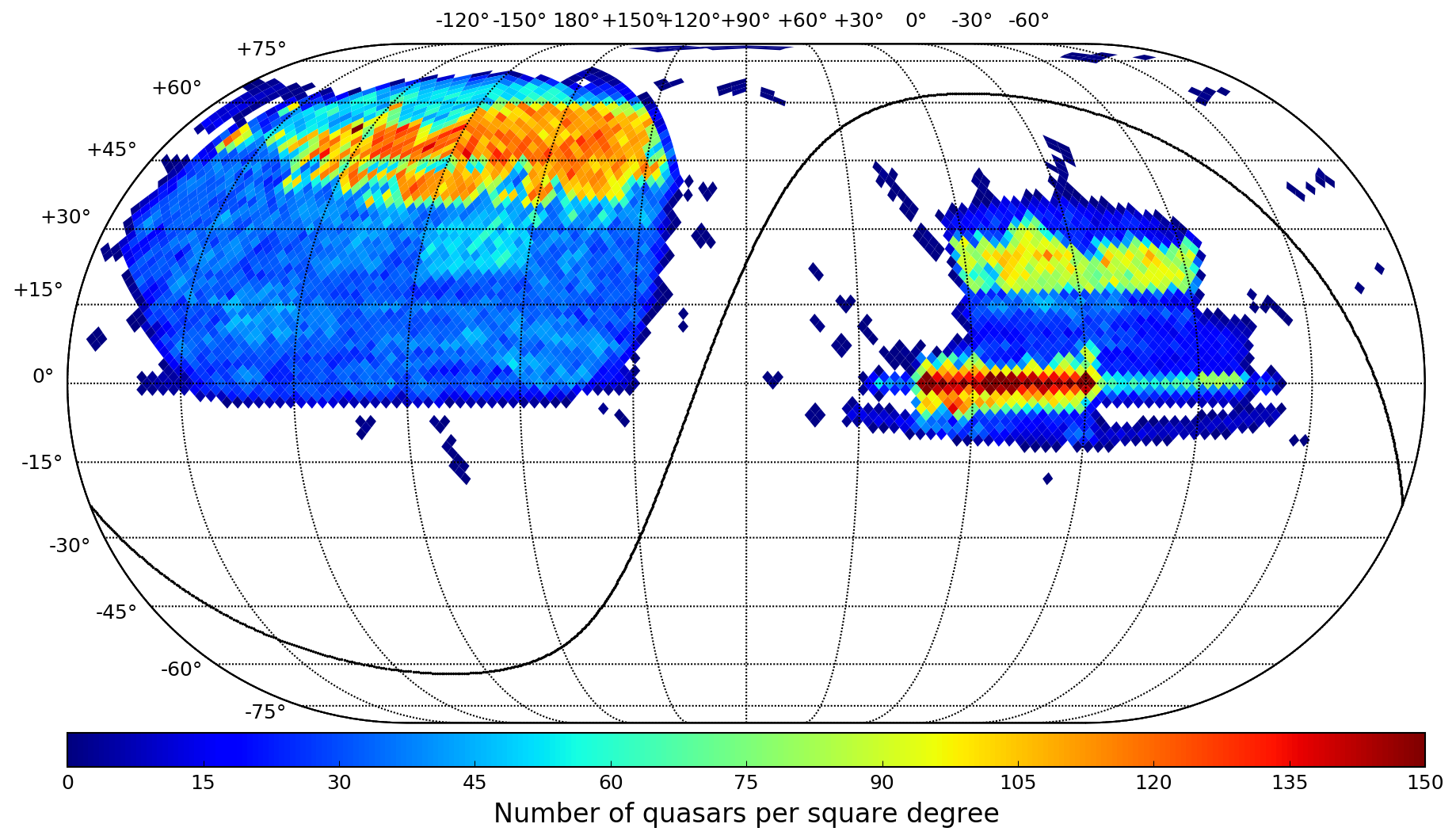}
   \caption{Sky distribution of the full DR14Q sample of 526,356 quasars in J2000 equatorial coordinates. The color coding indicates the number density of quasars in square degrees. The discontiguous small areas contain only SDSS DR7 quasars. The high-density regions are the eBOSS and SEQUELS observations.}
   \label{fig:sky}
   \end{center}
\end{figure}

The quasar sample used for this analysis was collected by different generations of the Sloan Digital Sky Survey \cite{2000AJ....120.1579Y} (SDSS) at Apache Point Observatory (APO). The bulk of the quasars was gathered over a five-year period by BOSS of SDSS-III. The major addition of new and reobserved quasars to the sample comes from the first two years of observations by eBOSS of SDSS-IV. These eBOSS spectra were made publicly available in DR14. The quasar target selection for eBOSS is described in \cite{2015ApJS..221...27M,2016A&A...587A..41P}. A smaller set of quasar spectra was contributed by the shorter program SEQUELS: The Sloan Extended QUasar, ELG and LRG Survey \cite{2015ApJS..221...27M}, designed as a pilot survey for eBOSS, which was undertaken as part of both SDSS-III and SDSS-IV. The sample also incorporates quasars from SDSS-I/II originally released in DR7 \cite{2010AJ....139.2360S}. The DR14 quasar catalog \cite{2017arXiv171205029P} (DR14Q) compiles the data from the aforementioned surveys and includes a total of 526,356 quasars (144,046 are new discoveries since the beginning of SDSS-IV) in the redshift range $0<z<7$. Figure~\ref{fig:sky} shows the quasar number density over the combined survey footprint.

All spectra used for this analysis were obtained using the BOSS spectrographs \cite{2013AJ....146...32S}, covering observed wavelengths $3600\lesssim\lambda\lesssim10,400~\angstrom$, with a wavelength-dependent resolving power $R\equiv \lambda/\Delta\lambda_{\rm FWHM}$ ranging from $\sim1300$ at the blue end of the spectrograph to $\sim2600$ at the red end. The data are reduced by the eBOSS pipeline, which performs wavelength calibration, flux calibration and sky subtraction of the spectra. The individual exposures of a given object are combined into a coadded spectrum that is rebinned onto pixels on a uniform grid with $\Delta\log_{10}(\lambda)=10^{-4}$ (corresponding to a velocity width $\Delta v\approx 69$~km~s$^{-1}$). In addition, the pipeline provides an automatic classification into object type (galaxy, quasar, star) and a redshift estimation by fitting a model spectrum \cite{2012AJ....144..144B}.

All SDSS-III spectra (BOSS and about half of the SEQUELS program) were visually inspected to correct for mis-classifications of object type and inaccurate redshift determinations by the pipeline \cite{2017A&A...597A..79P}. Starting in SDSS-IV, most of the objects are securely classified by the automated pipeline, with less than 10\% of the spectra requiring visual inspection \cite{2016AJ....151...44D}. DR14Q provides a column containing what is considered the most robust estimate of the redshift for each quasar. When available, the visual inspection redshifts are taken as the definitive quasar redshifts. Redshifts for quasars in SDSS-I/II are all based on visual inspection and taken from the DR7 quasar catalog \cite{2010AJ....139.2360S}. The remaining quasars have redshifts estimated by the pipeline.

The data are conveniently reorganised using the publicly available code \texttt{ForetFusion}\footnote{\url{https://github.com/igmhub/ForetFusion}} as follows. We split the sample into HEALPix pixels \cite{2005ApJ...622..759G} using a coarse pixelisation of the sky (HEALPix parameter \texttt{nside}=32). In each of these pixels, spectra with the same object identification \texttt{\detokenize{THING_ID}} are coadded using inverse variance as weights. Since the data reduction pipeline already places all spectral pixels on a fixed grid, no further interpolation is necessary, but the grid is extended to accommodate different start- and end-points in wavelength.

The cross-correlation analysis involves the selection of two quasar samples from DR14Q: tracer quasars, for which we only need the redshifts and positions on the sky, and forest quasars, for which we also use their spectra. We select 292,636 forest quasars in the redshift range $1.4<z_{\rm q}<3.5$ suitable for studying correlations from CIV absorption. The lower redshift limit is a consequence of the CIV forest exiting the wavelength coverage of the spectrograph for quasars with $z<1.4$. Because of the low number of quasars with $z>3.5$ in DR14Q, these quasars yield no significant contribution to this correlation study and are therefore not included. The forest sample excludes 33,497 quasars from SDSS DR7, as well as 21,030 broad absorption line (BAL) quasars (identified as having a balnicity index \cite{1991ApJ...373...23W} of the CIV absorption troughs \texttt{\detokenize{BI_CIV>}}0 in DR14Q). The selected sample of tracer quasars is larger, containing 387,315 quasars in the range $1.2<z_{\rm q}<3.5$, and includes the SDSS DR7 and BAL quasars.

We discard pixels which were flagged as problematic in the flux calibration or sky subtraction by the pipeline, and cut all pixels with observed wavelength $\lambda<3600~\angstrom$, below which the instrument throughput is less than 10\% of its peak value. In addition, we mask pixels around bright sky lines where the pipeline sky subtraction is found to be inaccurate. Pixels that satisfy the condition $\left| 10^{4}\log_{10}(\lambda/\lambda_{\rm sky})\right|\leq1.5$, where $\lambda_{\rm sky}$ is the wavelength at the pixel center of the sky line, are discarded. Lastly, we use a similar condition with double the margin to mask pixels around the observed CaII H\&K lines arising from absorption by the interstellar medium.

\begin{figure}
   \begin{center}
   \includegraphics[width=4.0in]{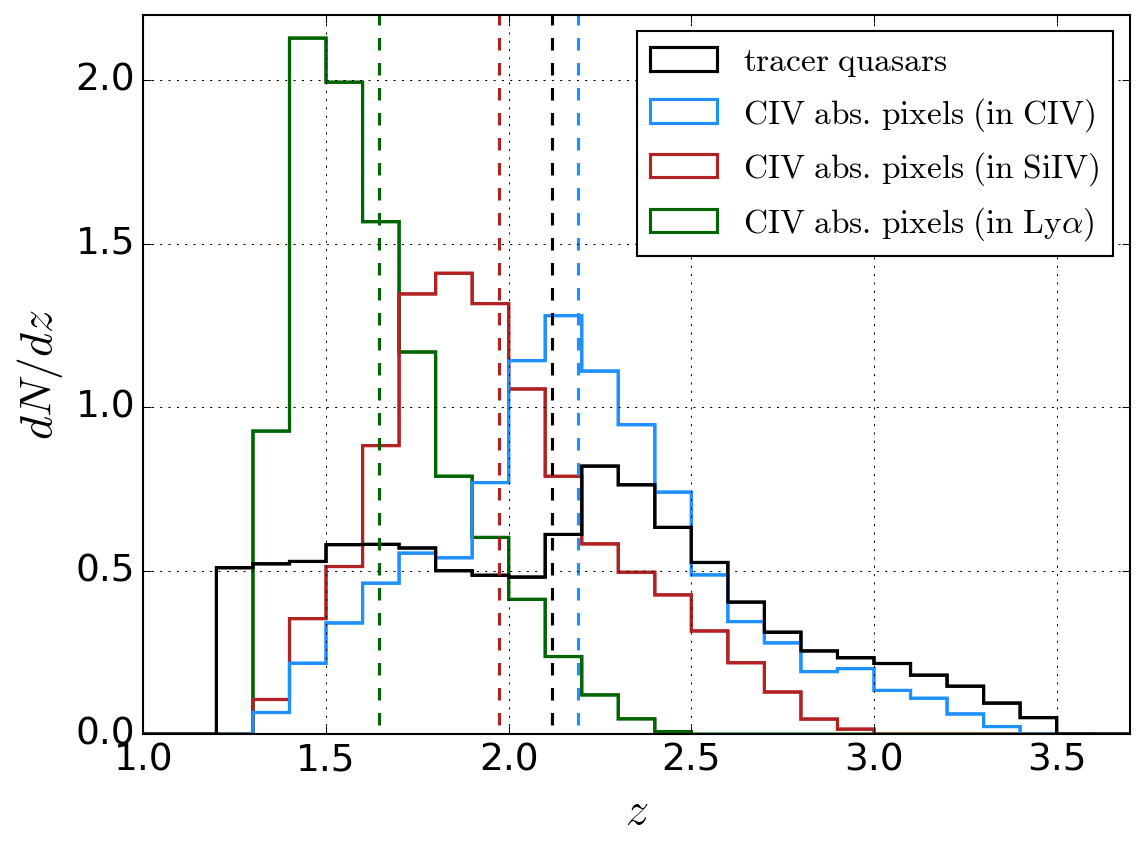}
   \caption{Normalized redshift distributions for the tracer quasars (black) and the CIV absorption pixels of the CIV forest (blue), the SiIV forest (red) and the \Lya forest (green). The absorption redshifts are calculated assuming CIV absorption at the rest-frame wavelength $1549.06~\angstrom$ and the pixels are weighted using the weights defined in equation~(\ref{eq:weights}). The histograms include 387,315 tracer quasars, $30.2\times10^{6}$ pixels in the CIV forest, $33.4\times10^{6}$ pixels in the SiIV forest, and $41.5\times10^{6}$ pixels in the \Lya forest. The vertical dashed lines show the mean value of each distribution: $\overline{z}= 2.12$ (tracer quasars), 2.19 (in CIV), 1.97 (in SiIV), 1.65 (in Ly$\alpha$).}
   \label{fig:zhisto}
   \end{center}
\end{figure}

Following the approach in \cite{2017A&A...603A..12B}, we define wider ``analysis pixels'' that are the inverse-variance-weighted flux average over three adjacent pipeline pixels. Requiring a minimum of 20 analysis pixels in each spectrum discards 3641 spectra. Finally, 1344 quasars failed the continuum-fitting procedure (see section~\ref{sec:transmission}). The total number of CIV forest quasars included in the final sample is 287,651. As discussed in section~\ref{subsec:cross}, in this work we also study correlations from CIV absorption occuring in the SiIV forest and the \Lya forest. In comparison, the subsamples used for the SiIV forest and the \Lya forest include 247,452 ($1.65<z<3.5$) and 188,665 ($2.05<z<3.5$) forest quasars, respectively. Figure~\ref{fig:zhisto} shows the redshift distributions for the tracer quasars and the CIV absorption pixels of the CIV forest, the SiIV forest and the \Lya forest.

For unsaturated lines, the relative line strength of the CIV doublet lines is deterministic (set by the ratio of the oscillator strengths) with the $1548.20~\angstrom$ line being twice as strong as the $1550.78~\angstrom$ line. The rest-frame wavelength separation of $\Delta\lambda_{\rm rf}=2.58~\angstrom$ between the doublet lines corresponds to a line-of-sight separation of about $5~\hMpc$ at $z=2$. This separation is small enough (only slightly larger than the width of the correlation function bins, $4~\hMpc$; see section~\ref{sec:crosscorrelation}) that distinguishing the individual contributions from the doublet lines  to the cross-correlation with quasars is challenging. We choose to employ the weighted mean wavelength, $\lambda_{\rm CIV}=1549.06~\angstrom$, as an effective CIV absorption line in this study. For the CIV forest region, we adopt the rest-frame wavelength interval
\begin{equation}
1420\,< \lambda_{\rm rf} <\,1520~\angstrom\ .
\end{equation}
This range is bracketed by the emission peaks of the SiIV doublet (with wavelengths $1393.76~\angstrom$ and $1402.77~\angstrom$) and the CIV doublet and was chosen as the maximum range that avoids the large pixel variances on the wings of the two emission peaks due to quasar-to-quasar diversity of line-emission strength. Similarly, we define the SiIV forest to be the rest-frame wavelength interval
\begin{equation}
1260\,< \lambda_{\rm rf} <\,1375~\angstrom\ .
\end{equation}
For the \Lya forest, we adopt the same range as used in \Lya BAO analyses \cite{2015A&A...574A..59D,2014JCAP...05..027F,2017A&A...603A..12B,2017A&A...608A.130D},
\begin{equation}
1040\,< \lambda_{\rm rf} <\,1200~\angstrom\ .
\end{equation}

The CIV forests collectively cover the observed wavelength range $3600<\lambda<6840~\angstrom$, corresponding to CIV absorption redshifts, $z_{\rm CIV}=\lambda/\lambda_{\rm CIV}-1$, in the range
\begin{equation}
1.32\,< z_{\rm CIV} <\,3.42\ ,
\end{equation}
as illustrated in figure~\ref{fig:zhisto}. CIV absorption in the SiIV forest is limited to $z_{\rm CIV}<2.99$ and in the \Lya forest to $z_{\rm CIV}<2.49$.

\section{Measurement of the flux-transmission field}
\label{sec:transmission}

The quantity of interest for performing correlation analyses using quasar absorption spectra is the transmitted flux fraction $F$, calculated in every pixel of the forest region of quasar $q$ as the ratio of the observed flux density $f_{\rm q}$ with the continuum flux $C_{\rm q}$ (the flux density that would be observed in the absense of absorption). The flux-transmission field $\delta_{\rm q}(\lambda)$ describes the fluctuations in the transmitted flux fraction relative to the mean value at the absorber redshift $\overline{F}(z)$:
\begin{equation}
\delta_{\rm q}(\lambda) = \frac{f_{\rm q}(\lambda)}{C_{\rm q}(\lambda)\overline{F}(z)}-1\ .
\label{eqn:deltafield}
\end{equation}

Following a similar approach to the one established for \Lya forest BAO analyses \cite{2013A&A...552A..96B,2015A&A...574A..59D}, we measure the flux-transmission field by estimating the product $C_{\rm q}(\lambda)\overline{F}(z)$ for each quasar. Each spectrum is modelled assuming a uniform forest spectral template which is multiplied by a linear function that sets the overall amplitude and slope:
\begin{equation}
C_{\rm q}(\lambda)\overline{F}(z) = \overline{f}(\lambda_{\rm rf})(a_{\rm q}+b_{\rm q}\log(\lambda))\ .
\end{equation}
The spectral template is derived from the data as a weighted mean normalized flux, obtained by stacking the spectra in the quasar rest frame.

\begin{figure}
   \begin{center}
   \includegraphics[width=4.0in]{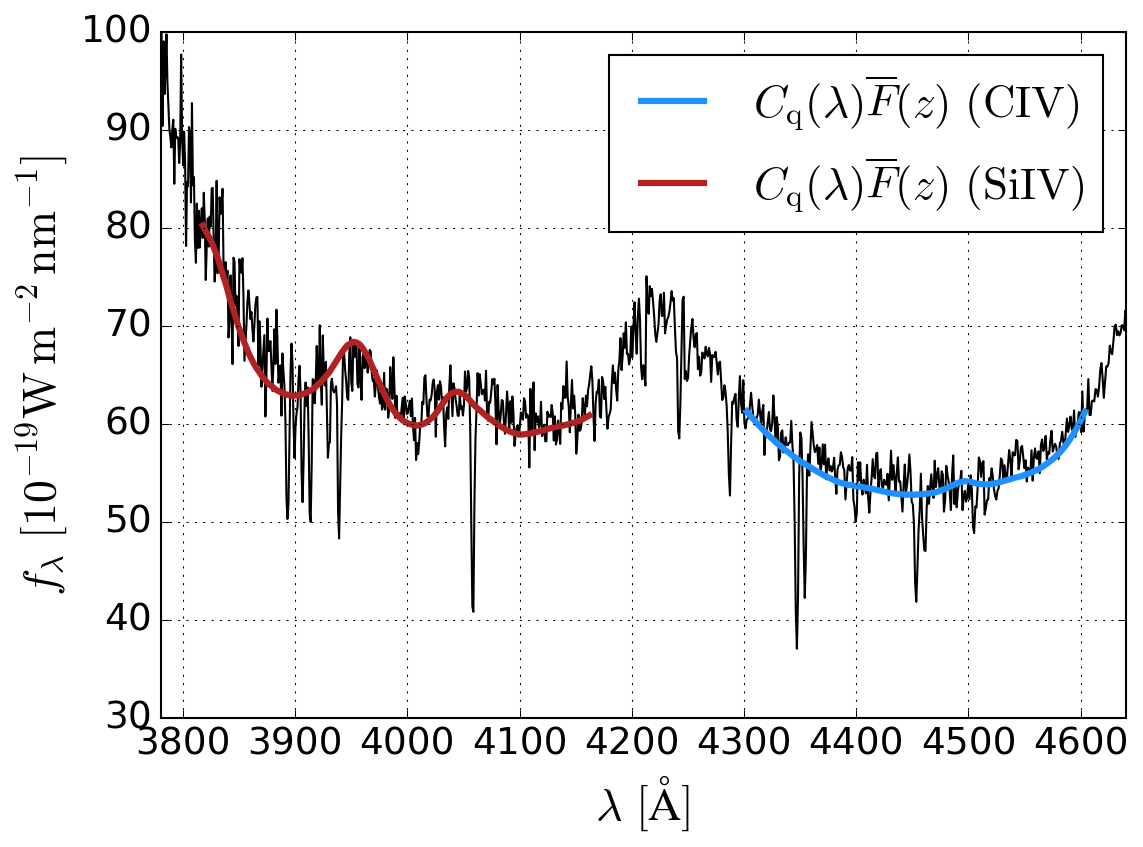}
   \caption{An example of a BOSS spectrum of a quasar at $z=2.03$ included in DR14Q. The spectrograph resolution at $\lambda\sim4000~\angstrom$ is $\sim2~\angstrom$. The blue line shows the best-fit model $C_{\rm q}(\lambda)\overline{F}(z)$ for the CIV forest covering the rest-frame wavelength interval $1420<\lambda_{\rm rf}<1520~\angstrom$ and the red line the SiIV forest covering $1260<\lambda_{\rm rf}<1375~\angstrom$.}
   \label{fig:spectrum}
   \end{center}
\end{figure}

We model the total variance of the delta field as
\begin{equation}
\sigma^{2} = \eta(z)\sigma_{\rm noise}^{2}+\sigma_{\rm LSS}^{2}(z)+\epsilon(z)/\sigma_{\rm noise}^{2}\ ,
\end{equation}
where $\sigma_{\rm noise}^{2}=\sigma_{\rm pipe}^{2}/(C_{\rm q}\overline{F})^{2}$. The first term represents the pipeline estimate of the flux variance, corrected by a function $\eta(z)$ that accounts for possible misestimation by the pipeline. The second term gives the contribution due to the large-scale structure, while the third term accounts for differences between the continuum template and the individual spectral shapes that become apparent at high signal-to-noise ratio. In bins of $\sigma_{\rm noise}^{2}$ and redshift, we measure the variance of the delta field and fit for the values of $\eta$, $\sigma_{\rm LSS}^{2}$ and $\epsilon$ as a function of redshift. The procedure of stacking the spectra, fitting the continua and measuring the variance of $\delta$ is iterated, until the three functions converge. Figure~\ref{fig:spectrum} displays an example spectrum together with the best-fit model $C_{\rm q}(\lambda)\overline{F}(z)$ for the CIV forest and the SiIV forest.

The pixel weights $w_{i}$ for the flux-transmission field are defined as the inverse of the total pixel variance:
\begin{equation}
w_{i} = \sigma_{i}^{-2}
\label{eq:weights}
\end{equation}

As described in \cite{2017A&A...603A..12B}, the delta field can be redefined in two steps to make explicit the biases introduced by the continuum fitting procedure. First, we define
\begin{equation}
\hat\delta_{\rm q}(\lambda) = \delta_{\rm q}(\lambda) - \overline{\delta_{\rm q}} - (\Lambda-\overline{\Lambda_{\rm q}})\frac{\overline{(\Lambda-\overline{\Lambda_{\rm q}})\delta_{\rm q}}}{\overline{(\Lambda-\overline{\Lambda_{\rm q}})^{2}}}
\quad , \quad
\Lambda\equiv\log(\lambda)\ ,
\end{equation}
where the over-bars refer to weighted averages over individual forests. In the second step, we transform the $\hat\delta_{\rm q}(\lambda)$ by subtracting the weighted average at each observed wavelength: $\hat\delta_{\rm q}(\lambda)\rightarrow\hat\delta_{\rm q}(\lambda)-\overline{\delta(\lambda)}$.

\section{The one-dimensional correlation function}
\label{sec:1Dcorr}

\begin{figure}
   \begin{center}
   \includegraphics[width=4.0in]{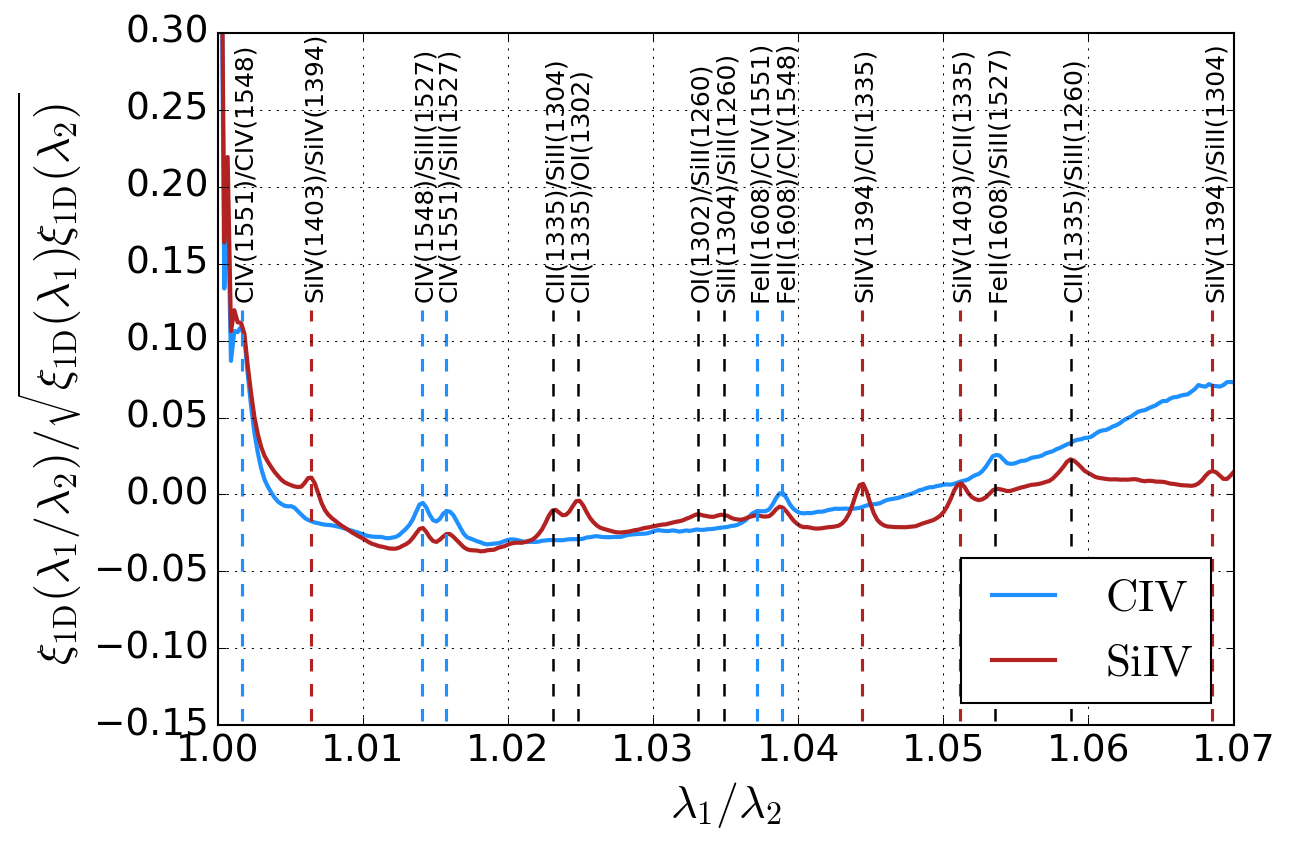}
   \caption{The one-dimensional correlation function, $\xi_{\rm 1D}$, as a function of wavelength ratio for pairs of pixels within the same forest (CIV or SiIV). Vertical, dashed lines indicate the prominent correlation peaks involving the CIV doublet lines (blue), the SiIV doublet lines (red) and other metal transitions (black).}
   \label{fig:corr1d}
   \end{center}
\end{figure}

Absorption in both the CIV forest and the SiIV forest are due to several transitions other than the CIV doublet. These contaminating absorptions by metals in the intergalactic medium may produce additional correlation signals in the cross-correlation with quasars that need to be modeled and marginalized out in the fit to the data. To identify the metal contaminants in each forest, we measure the one-dimensional correlation function, $\xi_{_{\rm 1D}}$, as the weighted mean of the product of $\delta_{\rm q}(\lambda)$ for pixel pairs within the same forest (see \cite{2017A&A...603A..12B,2017A&A...608A.130D} for similar studies of the contaminating absorptions in the \Lya forest). The pixel weights are defined in equation~(\ref{eq:weights}). Figure~\ref{fig:corr1d} shows $\xi_{_{\rm 1D}}$, normalized to unity at zero separation, as a function of wavelength ratio $\lambda_{1}/\lambda_{2}$ for the CIV forest and the SiIV forest. Prominent peaks are clearly visible from correlations between CIV doublet, CIV-metal, SiIV doublet, SiIV-metal and metal-metal pixel pairs.

The metal lines SiII $1526.71~\angstrom$ and FeII $1608.45~\angstrom$ yield correlations with the CIV doublet that are observed in both forests. Table~\ref{table:metals} summarizes the characteristics of these correlations. Both of these metal lines complicate the theoretical model
of the three-dimensional cross-correlation function by making the redshift of the absorption pixel ambiguous. We describe how to account for the quasar-metal cross-correlations in the model in section~\ref{sec:model}. For the SiIV forest, $\xi_{_{\rm 1D}}$ features additional peaks due to the SiIV doublet correlated with the metal lines SiII $1260.42~\angstrom$, OI $1302.17~\angstrom$, SiII $1304.37~\angstrom$ and CII $1334.53~\angstrom$ (as well as between the metal lines themselves). These absorptions due to transitions far from the CIV doublet transitions occur at widely separate redshifts and thus have negligible impact on the quasar-CIV cross-correlation.

\begin{table}
\begin{center}
\begin{tabular}{l c c c c}
\hline
\hline
\noalign{\smallskip}
Correlation & $\lambda_{1}~[\angstrom]$ & $\lambda_{2}~[\angstrom]$ & $\lambda_{1}/\lambda_{2}$ & $r_{\parallel}~[\hMpc]$ \\
\noalign{\smallskip}
\hline
\noalign{\smallskip}
CIV(1548)-SiII(1527) & 1548.20 & 1526.71 & 1.0141 & -41.5 \\
CIV(1551)-SiII(1527) & 1550.78 & 1526.71 & 1.0158 & -46.5 \\
FeII(1608)-CIV(1551) & 1608.45 & 1550.78 & 1.0372 & +109.6 \\
FeII(1608)-CIV(1548) & 1608.45 & 1548.20 & 1.0389 & +114.6 \\
\noalign{\smallskip}
\hline
\end{tabular}
\end{center}
\caption{Metal correlations involving the CIV doublet lines observed for pairs of pixels in the CIV forest and the SiIV forest. The second and third columns list the rest-frame wavelengths of the lines, and the fourth column their wavelength ratio. Line-of-sight separations are calculated at the observed wavelength $\lambda=4600~\angstrom$.}
\label{table:metals}
\end{table}

\section{The quasar-CIV forest cross-correlation}
\label{sec:crosscorrelation}

The three-dimensional positions of the sampled CIV transmission field and the quasars are defined by their redshift and angular coordinates on the sky (right ascension and declination). For every pair of quasar-CIV absorption, we transform the observed angular and redshift separations ($\Delta\theta,\Delta z$) into Cartesian coordinates ($r_{\perp},r_{\parallel}$) assuming a fiducial cosmology. The comoving separations along the line of sight $r_{\parallel}$ and transverse to the line of sight $r_{\perp}$ are defined as
\begin{equation}
r_{\parallel} = (D_{\rm CIV}-D_{\rm q})\cos\left(\frac{\Delta\theta}{2} \right)
\end{equation}
\begin{equation}
r_{\perp} = (D_{\rm CIV}+D_{\rm q})\sin\left(\frac{\Delta\theta}{2} \right)\ .
\end{equation}
Here, $D_{\rm CIV}\equiv D_{c}(z_{\rm CIV})$ and $D_{\rm q}\equiv D_{c}(z_{\rm q})$ are the comoving distances to the CIV absorption pixel and the quasar, respectively. The pair redshift is defined as $z_{\rm pair}=(z_{\rm CIV}+z_{\rm q})/2$.

The fiducial cosmology used for the analysis of the data is a flat $\Lambda$CDM model with parameter values taken from the Planck 2015 result (using the TT+lowP combination) \cite{2016A&A...594A..13P}: $\Omega_{\rm c}h^{2}=0.1197$, $\Omega_{\rm b}h^{2}=0.0222$, $\Omega_{\nu}h^{2}=0.0006$, $h=0.6731$, $N_{\nu}=3$, $\sigma_{8}=0.8298$, $n_{s}=0.9655$, $\Omega_{m}=0.3146$. The sound horizon at the drag epoch is calculated using CAMB \cite{2000ApJ...538..473L}, $r_{d}=99.17~\hMpc$.

\subsection{Cross-correlation}
\label{subsec:cross}

We estimate the cross-correlation at a separation bin $A$, $\xi_{A}$, as the weighted mean of the $\delta$ field in pairs of pixel $i$ and quasar $k$ at a separation within the bin $A$ (see \cite{2012JCAP...11..059F} for a formal derivation of this expression and \cite{2017A&A...608A.130D} for tests on simulated data):
\begin{equation}
\xi_{A}=\frac{\sum\limits_{(i,k)\in A}w_{i}\delta_{i}}{\sum\limits_{(i,k)\in A}w_{i}}\ ,
\end{equation}
where the weights $w_{i}$, defined in equation~(\ref{eq:weights}), take into account both instrumental noise and small-scale absorption fluctuations.

Our separation grid consists of 100 bins of $4~\hMpc$ within $-200~\hMpc$ and $+200~\hMpc$ in the parallel direction and 50 bins of $4~\hMpc$ within $0~\hMpc$ and $+200~\hMpc$ in the perpendicular direction, for a total of 5000 bins. Each bin is defined by the weighted mean ($r_{\perp},r_{\parallel}$) of the quasar-pixel pairs of that bin, and its redshift by the weighted mean pair redshift. The effective redshift of the cross-correlation measurement is defined to be the inverse-variance-weighted mean of the redshift bins with separations in the BAO region $80<r<120~\hMpc$.

For the measurements of the cross-correlation, we use CIV absorptions in the CIV forest, the SiIV forest and the \Lya forest. We do not attempt to distiguish SiIV and \Lya absorptions from CIV absorptions in their respective forest as this would be a daunting task at eBOSS resolution and signal-to-noise ratio levels. The SiIV and \Lya transitions are sufficiently separated in wavelength from the CIV transition, respectively corresponding to large physical separations of $\sim300~\hMpc$ and $\sim700~\hMpc$, to neglect the contamination of SiIV and \Lya absorption interpreted as CIV absorption (i.e., $b_{\rm CIV}\xi_{\rm L}(r=10~\hMpc)
>>b_{\rm Ly\alpha}\xi_{\rm L}(r=700~\hMpc)$, where $\xi_{\rm L}$ is the linear matter correlation function). All SiIV and \Lya absorptions then act purely as a source of additional variance in the measurement. There are $10.4\times10^{9}$ quasar-pixel pairs included in the cross-correlation measurement for the CIV forest. In comparison, the measurement for the SiIV forest includes $10.8\times10^{9}$ pairs and the \Lya forest $9.9\times10^{9}$ pairs.

\subsection{Distortion matrix}
\label{subsec:distortion}

The procedure used to estimate the delta field, described in section~\ref{sec:transmission}, suppresses fluctuations of characteristic scales corresponding to the length of a forest, since the estimate of $C\overline{F}$ in equation~(\ref{eqn:deltafield}) would follow such a fluctuation. This results in a suppression of power in the radial direction and a significant distortion in the correlation function. The effect of this suppression on the power spectrum was studied in \cite{2015JCAP...11..034B} where a sigmoidal-like factor, a function of $k_{\parallel}$ (the wave-number along the line of sight), was introduced and optimized using simulations.

In this paper, we follow instead the approach introduced by \cite{2017A&A...603A..12B} and adapted to the cross-correlation by \cite{2017A&A...608A.130D} which allows one to encode the effect of this distortion on the correlation function in a ``distortion matrix''. This approach was extensively validated in these references using simulated data. The observed correlation function is the product of this matrix and the expected correlation function:
\begin{equation}
\hat\xi_{A} = \sum\limits_{A^\prime}D_{AA^\prime}\xi_{ A^\prime}\ .
\end{equation}
The quantity $D_{AA^\prime}$ is
\begin{equation}
D_{AA^\prime} = \frac{\sum\limits_{(i,k)\in A}w_{i}\sum\limits_{(j,k^{\prime})\in A^{\prime}}\left(\delta_{ij}^{K}-\frac{w_{j}}{\sum\limits_{l}w_{l}}-\frac{w_{j}(\Lambda_{i}-\overline{\Lambda})(\Lambda_{j}-\overline{\Lambda})}{\sum\limits_{l}w_{l}(\Lambda_{l}-\overline{\Lambda})^{2}}\right)}{\sum\limits_{(i,k)\in A}w_{i}} \quad , \quad
\Lambda\equiv\log(\lambda)\ ,
\end{equation}
where the sums run over all delta-field pixel ($i,j$) and quasar ($k, k^{\prime}$) pairs that contribute to the separation bins $A$ and $A'$, and $\delta^K$ is the Kronecker delta. We use this distortion matrix when comparing models for the correlation function to our measurements (see section~\ref{sec:model}) and performing fits to the data.

\subsection{Covariance matrix}
\label{subsec:covariance}

We calculate the covariance matrix by using the subsampling technique introduced by \cite{2013A&A...552A..96B} and adapted to the cross-correlation by \cite{2017A&A...608A.130D}. We divide the space of quasar-pixel pairs in subsamples and measure the covariance from the variability across the subsamples. Such estimates of the covariance matrix are unbiased, but the noise due to the finite number of subsamples leads to biases in the inverse of the covariance \cite{2014IAUS..306...99J}. Following \cite{2017A&A...608A.130D}, we smooth the noise by assuming that the covariance between separation bins $A$ and $B$ depends only on the absolute difference ($|r_\parallel^A-r_\parallel^B|$, $|r_\perp^A-r_\perp^B|$).

The subsamples are defined as follows. We use HEALPix to define a pixelization of the sky. A quasar-absorption pixel pair is assigned to a subsample $s$ if the forest that contains the absorption belongs to the HEALPix pixel $s$.

\begin{figure}
   \begin{center}
   \includegraphics[width=3.0in]{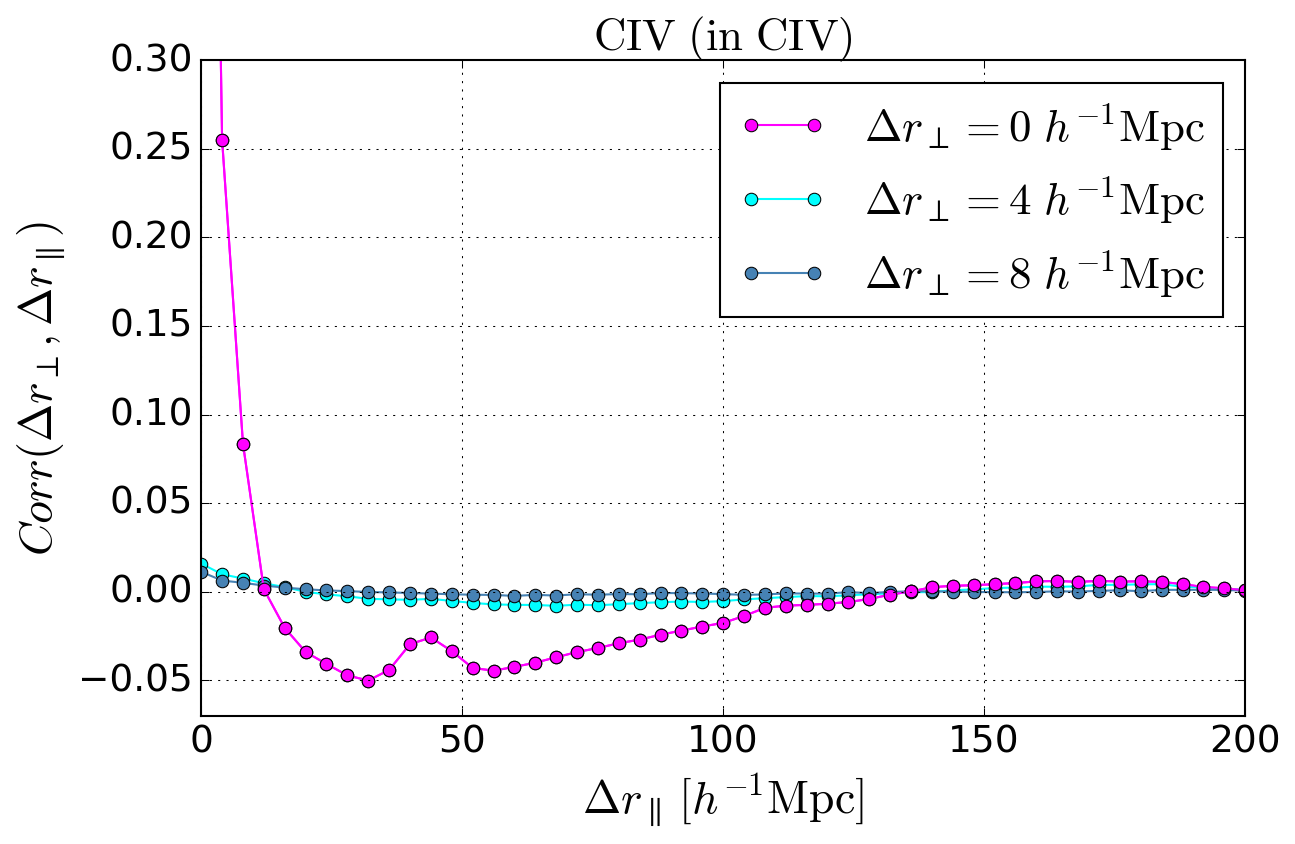}
   \includegraphics[width=3.0in]{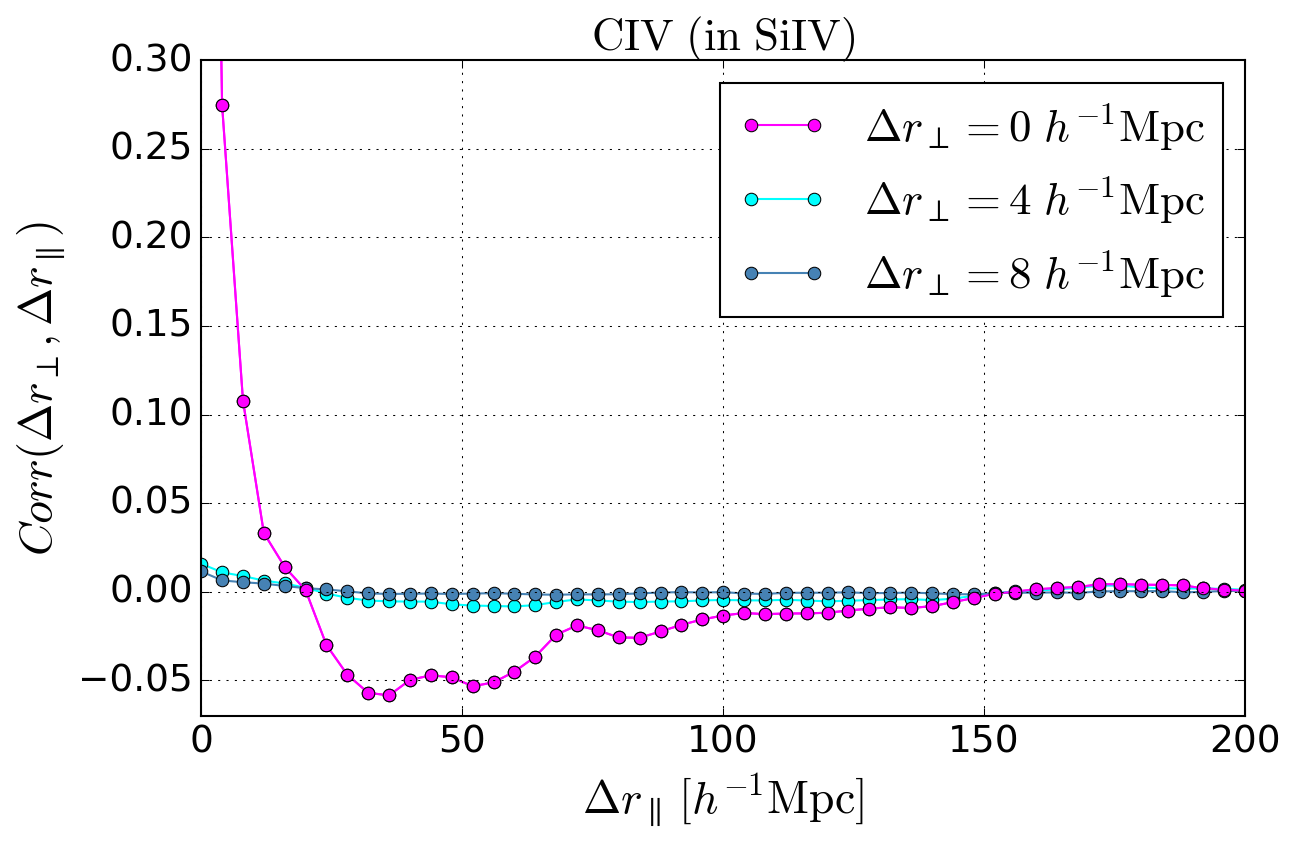}
   \caption{The mean correlation matrix, $Corr_{AB}$, as a function of $\Delta r_{\parallel}=|r_{\parallel,A}-r_{\parallel,B}|$ for CIV absorption in the CIV forest (left) and the SiIV forest (right) determined by subsampling. The curves are for constant $\Delta r_{\perp}=|r_{\perp,A}-r_{\perp,B}|$ for the three lowest values $\Delta r_{\perp}=\left[0,4,8\right]~\hMpc$. Correlations for values of $\Delta r_{\parallel}$ beyond the plotted range ($\Delta r_{\parallel}>200~\hMpc$) are approximately zero.}
   \label{fig:corrmat}
   \end{center}
\end{figure}

Before the smoothing procedure, our ``raw'' covariance is calculated as follows:
\begin{equation}
C^\mathrm{raw}_{AB} = \frac{1}{W_{A}W_{B}}\sum\limits_{s}W_{A}^{s}W_{B}^{s}\left[\xi_{A}^{s}\xi_{B}^{s}-\xi_{A}\xi_{B}\right]\ ,
\end{equation}
where $W_{A}$ is the sum of the pair weights $w$ belonging to bin $A$,
\begin{equation}
W_{A}=\sum\limits_{i\in A}w_{i}\ .
\end{equation}
From this raw covariance, we calculate the raw correlation matrix:
\begin{equation}
Corr^\mathrm{raw}_{AB}=\frac{C_{AB}}{\sqrt{C_{AA}C_{BB}}}\ .
\end{equation}
Our smoothing procedure is then applied to this raw correlation matrix. The final covariance used in the fits is obtained by multiplying the smoothed correlation matrix by the diagonal of the raw covariance matrix. Some elements of the smoothed correlation matrix are shown in figure~\ref{fig:corrmat}.

\section{Model of the cross-correlation}
\label{sec:model}

Our model of the cross-correlation is an adaptation of the model employed for the cross-correlation of quasars with \Lya forest absorption \cite{2017A&A...608A.130D}. CIV absorbers in the intergalactic medium constitute the dominant contributors to the correlation signal obtained from the CIV forest absorption, but additional contributions arise from other metal absorbers with similar rest-frame wavelengths to the CIV doublet. As a result, the expected value of the measured cross-correlation in the bin $A$ is the sum of the cosmological quasar-CIV cross-correlation and quasar-metal cross-correlations, modified by the distortion matrix to correct for the distortions introduced by the continuum fitting procedure,
\begin{equation}
\hat\xi_{A} = \sum\limits_{A^\prime}D_{AA^\prime}\left[\xi_{{\rm qCIV}, A^\prime}+\sum\limits_{m}\xi_{{\rm qm},A^\prime}\right]\ .
\label{eqn:ximeasured}
\end{equation}

The cosmological quasar-CIV cross-correlation function is assumed to be a biased version of the total matter correlation function of the fiducial cosmological model. We adopt the standard methodology of decoupling the cross-correlation function into a smooth component and a peak component, to free the position of the BAO peak,
\begin{equation}
\xi_{\rm qCIV}(r_\parallel,r_\perp,\alpha) = \xi_{\rm smooth} (r_\parallel, r_\perp) + A_{\rm peak} \cdot \xi_{\rm peak} (r_\parallel,r_\perp,\alpha)\ .
\end{equation}
Here, the parameter $A_{\rm peak}$ controls the BAO peak amplitude and is kept fixed to its nominal value $A_{\rm peak}=1$. Due to the low signal-to-noise ratio of the current cross-correlation measurement, we limit the determination of the BAO peak position to the isotropic signal. The shift of the observed peak position relative to the fiducial peak position is thus described by the isotropic scale parameter
\begin{equation}
\alpha = \frac { \left[D_{V}(z_{\rm eff})/r_d\right] }{\left[D_{V}(z_{\rm eff})/r_{d}\right]_{\rm fid}}\ ,
\end{equation}
where $z_{\rm eff}$ is the effective redshift of the BAO measurement. The spherically-averaged distance $D_{V}$ constrained in the BAO fit is a combination of the BAO scale in the radial and transverse directions, governed by the Hubble parameter $H(z)$ and the comoving angular diameter distance $D_{M}(z)$, respectively,
\begin{equation}
D_{V}(z) = \left[czH^{-1}(z)D_{M}^{2}(z) \right]^{1/3}\ .
\end{equation}
In the case of \Lya correlations, the large effect of redshift-space distortions changes the relative contributions of the radial and transverse signal, making the compression into $D_{V}$ invalid. For CIV correlations, however, we expect the strength of redshift-space distortions to be significantly smaller (since metals are at least partially associated with the extended environments of galaxies) therefore the expression for $D_{V}$ will be approximately correct. This choice is born out by our subsequent measurement of redshift-space distortions (see figure~\ref{fig:civbeta}).

We model the cosmological quasar-CIV cross-correlation function as the Fourier transform of the cross-power spectrum:
\begin{equation}
P_{\rm qCIV}(\bm{k},z) = b_{\rm q}(z)\left[1 + \beta_{\rm q}\mu_{k}^{2}\right]b_{\rm CIV}(z)\left[1 + \beta_{\rm CIV}\mu_{k}^{2}\right]P_{\rm QL}(\bm{k},z)\sqrt{V_{\rm NL}(k_{\parallel})}G(\bm{k})\ ,
\end{equation}
where $\bm{k}=(k_{\parallel}, k_{\perp})$ is the wavenumber of modulus $k$ and $\mu_{k}=k_{\parallel}/k$ is the cosine of the angle between the wavenumber and the line of sight. The model builds on the linear theory prediction \cite{1987MNRAS.227....1K}, involving the standard factors of linear bias $b_{\rm CIV}(z)$ and redshift-space distortion $\beta_{\rm CIV}$ for the CIV absorption, and correspondingly $b_{\rm q}(z)$ and $\beta_{\rm q}$ for the quasars. Extensions to non-linear theory are introduced through the models of the quasi-linear matter power spectrum $P_{\rm QL}$, the non-linear quasar velocities $V_{\rm NL}$ and the effect of the binning on the measurement $G(\bm{k})$. These functions are described in more detail below.

Decoupling of the peak from the smooth component of the correlation function requires a corresponding separation of the (quasi) linear matter power spectrum:
\begin{equation}
P_{\rm QL}(\bm{k},z) =P_{\rm smooth}(k,z) + \exp\left[-(k_{\parallel}^2\Sigma_{\parallel}^2+k_{\perp}^2\Sigma_{\perp}^2)/2\right]P_{\rm peak}(k,z)\ .
\label{eqn:pql}
\end{equation}
Here, the smooth component $P_{\rm smooth}$ is derived from the linear matter power spectrum $P_{\rm L}$ by erasing the BAO peak following the sideband method of \cite{2013JCAP...03..024K}, and the peak component $P_{\rm peak}=P_{\rm L}-P_{\rm smooth}$. The linear matter power spectrum is obtained from CAMB at the effective redshift $z_{\rm eff}$ using the assumed fiducial cosmology. The exponential function in equation~(\ref{eqn:pql}) models anisotropic non-linear broadening of the BAO peak caused by the effect of large-scale bulk velocity flows on the matter density field \cite{2007ApJ...664..675E}. The non-linear broadening parameters are related by
\begin{equation}
\frac{\Sigma_{\parallel}}{\Sigma_{\perp}}=1+f\ ,
\end{equation}
where
\begin{equation}
f=d(\ln g)/d(\ln a)\approx\Omega_{\rm m}^{0.55}(z)
\end{equation}
is the linear growth rate of structure \cite{2007APh....28..481L} and $g(z)$ is the growth factor. We adopt the value $\Sigma_{\perp}=3.26~\hMpc$ \cite{2013JCAP...03..024K,2007ApJ...664..660E}, which together with the value of $f$ for the fiducial cosmology determines the value of $\Sigma_{\parallel}$. The redshift evolution of the linear matter power spectrum is governed by the growth factor such that (for $f\approx1$)
\begin{equation}
P_{\rm L}(k,z) = P_{\rm L}(k,z_{\rm eff})g^{2}(z)=P_{\rm L}(k,z_{\rm eff})\left( \frac{1+z_{\rm eff}}{1+z} \right)^{2}\ .
\end{equation}

The function $V_{\rm NL}(k_{\parallel})$ is used to model the effect on the power spectrum of non-linear velocities of the quasars and the dispersion of the quasar redshift measurements. It takes the functional form of a Lorentz damping factor \cite{2009MNRAS.393..297P},
\begin{equation}
V_{\rm NL}(k_{\parallel}) = \frac{1}{1+(k_{\parallel}\sigma_{v})^2}\ ,
\end{equation}
where $\sigma_{v}$ is a free parameter given in the unit $\hMpc$.

Following \cite{2017A&A...603A..12B}, the function $G(\bm{k})$ corrects for the smoothing of the correlation function due to the binning in $(r_{\perp},r_{\parallel})$ space,
\begin{equation}
G(\bm{k}) = \left[\mathrm{sinc}\left(\frac{k_{\parallel}R_{\parallel}}{2}\right)\mathrm{sinc}\left(\frac{k_{\perp}R_{\perp}}{2}\right)\right]^2\ ,
\end{equation}
where $R_{\parallel}$ and $R_{\perp}$ are the smoothing scales. These parameters are kept fixed to the bin width, $R_{\parallel}=R_{\perp}=4~\hMpc$. The functional form of $G(\bm{k})$ is an approximation for the transverse direction that greatly simplifies the correlation function computation and produces a sufficiently accurate result, as discussed in \cite{2017A&A...603A..12B}.

The linear bias of CIV absorption is assumed to evolve as a power law with redshift,
\begin{equation}
b_{\rm CIV}(z) = b_{\rm CIV}(z_{\rm eff})\left( \frac{1+z}{1+z_{\rm eff}} \right)^{\gamma}\ .
\label{eq:biasevo}
\end{equation}
We set $\gamma = 1$, which is a weaker redshift-evolution than that observed for measurements of the \Lya forest flux correlations ($\gamma_{\rm Ly\alpha} = 2.9$) \cite{2006ApJS..163...80M}. As presented in section~\ref{sec:results}, a linear evolution with redshift is consistent with the results obtained when comparing the fits after splitting the data into two redshift samples. The fit is not sensitive to the precise value assumed for $\gamma$ because of the narrow range of mean redshifts of the data bins. To simplify the fitting procedure, we assume that the redshift-space distortion parameter $\beta_{\rm CIV}$ is redshift independent across this range.

Because of the degeneracy between $b_{\rm CIV}\equiv b_{\rm CIV}(z_{\rm eff})$ and the quasar bias $b_{\rm q}\equiv b_{\rm q}(z_{\rm eff})$ in the fit, for each measurement we fix the value of $b_{\rm q}$ and assume a redshift-dependence across the bins given by \cite{2005MNRAS.356..415C,2017JCAP...07..017L}
\begin{equation}
b_{\rm q}(z) = 0.53 + 0.289(1+z)^{2}\ .
\end{equation}
For quasars, the redshift-space distortion parameter $\beta_{\rm q}$ is determined by the product
\begin{equation}
b_{\rm q}\beta_{\rm q} = f = \Omega_{\rm m}^{0.55}(z_{\rm eff})\ .
\end{equation}

The contribution from the quasar-metal cross-correlations in equation~(\ref{eqn:ximeasured}) is comparatively small and therefore treated in a simplified approach in which the fiducial correlation model from linear theory is used without separating it into smooth and peak components. The relevant metal contaminants for CIV absorption were identified in section~\ref{sec:1Dcorr} to be the transitions SiII $1526.71~\angstrom$ and FeII $1608.45~\angstrom$. Each metal component is described by its own linear bias and redshift-space distortion parameter ($b_{\rm m},\beta_{\rm m}$).

The redshift  $z_\mathrm{m}$ of these contaminant absorptions will be different from the assigned redshift $z_\mathrm{CIV}$ obtained assuming that all absorptions correspond to CIV absorption. Consequently, the fiducial distance to these absorptions will be different from the fiducial distance to the assumed CIV absorption by the amount
\begin{equation}
D_{\rm m} - D_{\rm CIV} = c\int_{z_\mathrm{CIV}}^{z_\mathrm{m}}\frac{dz^{\prime}}{H(z^{\prime})}\simeq \frac{c(1+\overline{z})}{H(\overline{z})}\frac{(\lambda_{\rm CIV}-\lambda_{\rm m})}{\overline{\lambda}}\ ,
\label{eqn:rcomov_shift}
\end{equation}
where $\overline{z} = (z_{\rm CIV}+z_{\rm m})/2$ and $\overline{\lambda} = (\lambda_{\rm CIV}+\lambda_{\rm m})/2$. The result of this is a shift along the line of sight for the quasar-metal cross-correlation. These metal contaminations are most visible when the physical separation between the quasar and the absorber is small, i.e., when $r_{\perp}\sim 0$ and $r_\parallel\sim -(1+\overline{z})(c/H(\overline{z}))(\lambda_{\rm CIV}-\lambda_{\rm m})/\overline{\lambda}$.

We calculate the metal contributions in equation~(\ref{eqn:ximeasured}) directly over the geometry of the survey in the following manner. For a given quasar-metal cross-correlation, $\xi_{\rm qm}$,
\begin{equation}
\xi_{{\rm qm}, A} = \frac{1}{W_{A}}\sum_{(i,k)\in A}w_{i}\xi_{\rm qm}(r^{ik}_{\parallel}, r^{ik}_{\perp})\ ,
\label{eqn:xiqm}
\end{equation}
where the sum runs over the pixel-quasar pairs $(i,k)$ within the separation bin $A$, using $z_{\mathrm{CIV}}$ to calculate the fiducial comoving distance to the pixel, but using $z_{\mathrm{m}}$ to calculate $r^{ik}_{\parallel}$ and $r^{ik}_{\perp}$. $W_{A}$ is the sum of weights in bin $A$. $\xi_{\rm qm}$ can be written in terms of a pixelization: $\xi_{\rm qm}(r^{ik}_{\parallel}, r^{ik}_{\perp})=\sum_{(i,k)\in B}\xi_{{\rm qm,} B}$, where $B$ represents a separation bin. Inserting this expression in equation~(\ref{eqn:xiqm}) yields
\begin{equation}
\xi_{{\rm qm}, A} = \frac{1}{W_{A}}\sum_B\xi_{{\rm qm},B}\sum_{(i,k)\in A, (i,k)\in B}w_{i} \equiv \sum_{B} M_{AB} \xi_{{\rm qm},B}\ ,
\end{equation}
where
\begin{equation}
M_{AB} \equiv \frac{1}{W_{A}}\sum_{(i,k)\in A, (i,k)\in B}w_{i}
\end{equation}
is a ``metal distortion matrix'' that allows us to calculate the shifted quasar-metal cross-correlation function for a given non-shifted quasar-metal cross-correlation function. The condition $(i,k)\in A$ refers to pixel distances calculated using $z_{\mathrm{CIV}}$, but $(i,k)\in B$ refers to pixel distances calculated using $z_{\mathrm{m}}$.

Because the amplitudes of the quasar-metal correlations are mostly determined by the excess correlation near the line of sight ($r_{\perp}\sim0$), the redshift-space distortion parameter for each metal is not well-constrained in the fit. Following \cite{2017A&A...608A.130D}, we therefore fix their value to $\beta_{\rm m} = 0.5$. The fit results are not sensitive to this choice. As for $b_{\rm CIV}(z)$, we assume a linear redshift evolution for the metal biases $b_{\rm m}(z)$.

Systematic errors in the quasar redshift estimates lead to a shift of the cross-correlation along the line of sight,
\begin{equation}
\Delta r_{\parallel,\rm q} = r_{\parallel, {\rm true}} - r_{\parallel, {\rm measured}} = \frac{(1+z)\Delta v_{\parallel}}{H(z)}\ .
\end{equation}
The radial coordinate shift, $\Delta r_{\parallel,\rm q}$, is a free parameter that may change with the mean redshift of the quasar sample due to different emission lines entering or leaving the wavelength coverage of the spectrograph and their different relative velocities due to their association with outflows \cite{1982ApJ...263...79G,2016ApJ...831....7S}.

While $\Delta r_{\parallel,\rm q}$ provides a mean shift of the cross-correlation along the line of sight, asymmetry between the foreground and background is introduced by the continuum-fitting distortion (described by the distortion matrix), the metal absorption, and the variation of the mean redshifts with $r_{\parallel}$. Additional asymmetry from relativistic effects has been considered in other studies \cite{2014PhRvD..89h3535B,2016JCAP...02..051I} but are not included here.

The low signal-to-noise ratio of the current cross-correlation measurement implies that the errors on the free parameters of the model are dominated by the statistical uncertainties. We therefore choose to not include a smoothly-varying ``broadband function'' which is normally introduced in \Lya forest BAO fitting to test for possible systematic errors in the determination of the BAO peak position.

Table~\ref{table:parameters} summarizes the parameters of the model. There are seven free parameters in the fits.

\begin{table}
\begin{center}
\begin{tabular}{ll}
\hline
\hline
\noalign{\smallskip}
Parameter & Description\\
\noalign{\smallskip}
\hline
\noalign{\smallskip}
$\beta_{\rm CIV}$ & Redshift-space distortion parameter for CIV absorption\\
$b_{\rm CIV}(1+\beta_{\rm CIV})$ & Combined bias parameters for CIV absorption\\
$\alpha$ & Isotropic BAO peak-position parameter\\
$\sigma_{v}$ & Quasar radial velocity smearing\\
$\Delta r_{\parallel,\rm q}$ & Coordinate shift due to the quasar redshift systematic error\\
$b_{\rm m}$ & Bias parameters of two metal species\\
\noalign{\smallskip}
\hline
\noalign{\smallskip}
$b_{\rm q}$ & Quasar bias parameter\\
$f$ & Linear growth rate of structure\\
$\gamma=1$ & Redshift evolution exponent for the bias of absorbers\\
$\Sigma_{\perp}=3.26~\hMpc$ & Transverse non-linear broadening of the BAO peak\\
$R_{\parallel}=4~\hMpc$ & Radial binning smoothing parameter\\
$R_{\perp}=4~\hMpc$ & Transverse binning smoothing parameter\\
$A_{\rm peak}=1$ & BAO peak amplitude\\
$\beta_{\rm m}=0.5$ & Redshift-space distortion parameters of two metal species\\
\noalign{\smallskip}
\hline
\end{tabular}
\end{center}
\caption{List of the parameters of the cross-correlation model. The fit parameters are given in the first part of the table. The second section lists parameters that are fixed in the fits.}
\label{table:parameters}
\end{table}

\section{Results}
\label{sec:results}

\begin{figure}
   \begin{center}
   \includegraphics[width=4.0in]{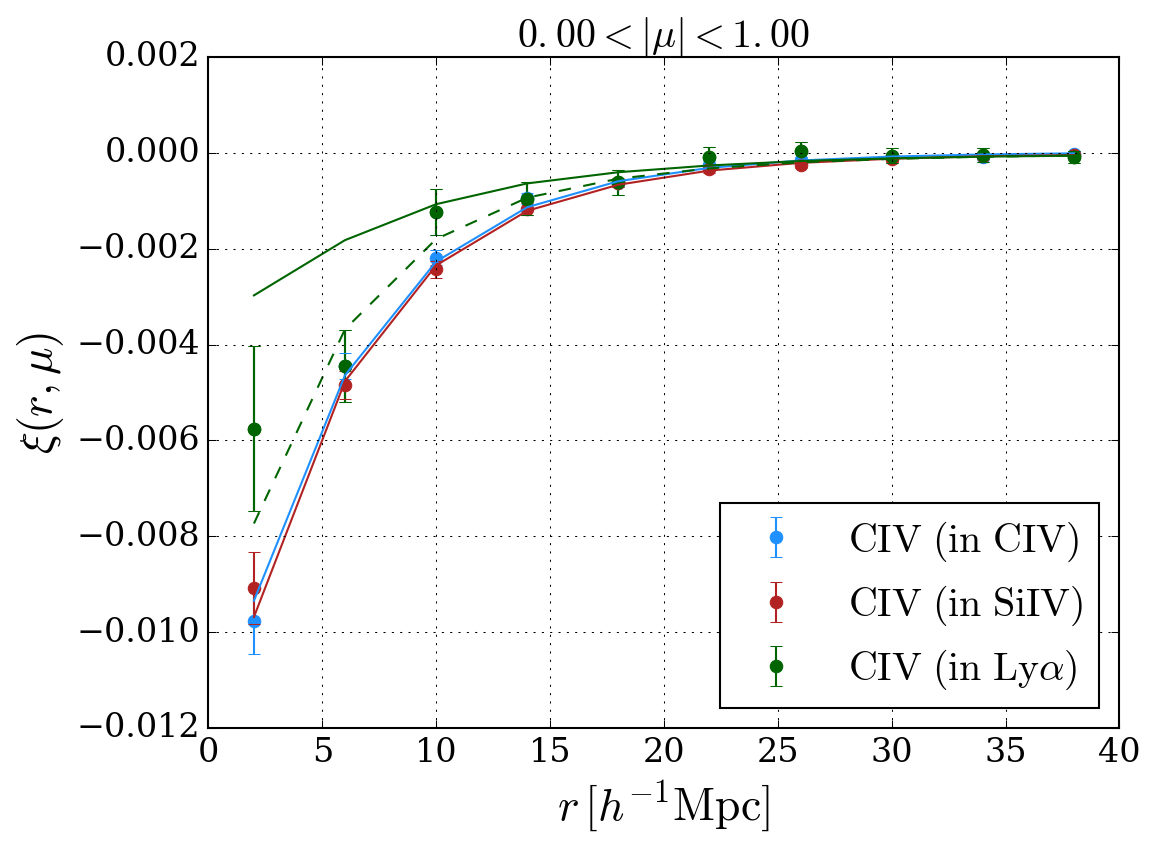}
   \caption{Cross-correlation functions averaged over the full angular range $0<|\mu|<1$ for CIV absorption measured in the CIV forest (blue), the SiIV forest (red) and the Ly$\alpha$ forest (green). The solid lines show the best-fit models obtained for the fitting range $10<r<180~\hMpc$ and extrapolated in the plot to separations $r<10~\hMpc$. The dashed green line shows the best-fit model for CIV (in Ly$\alpha$) for the fitting range $5<r<180~\hMpc$ (see the text and table~\ref{table:bestfitsrmin5}). The effective redshift is different for different forests.}
   \label{fig:civsiivlyawedge}
   \end{center}
\end{figure}

\begin{figure}
   \begin{center}
   \includegraphics[width=3.0in]{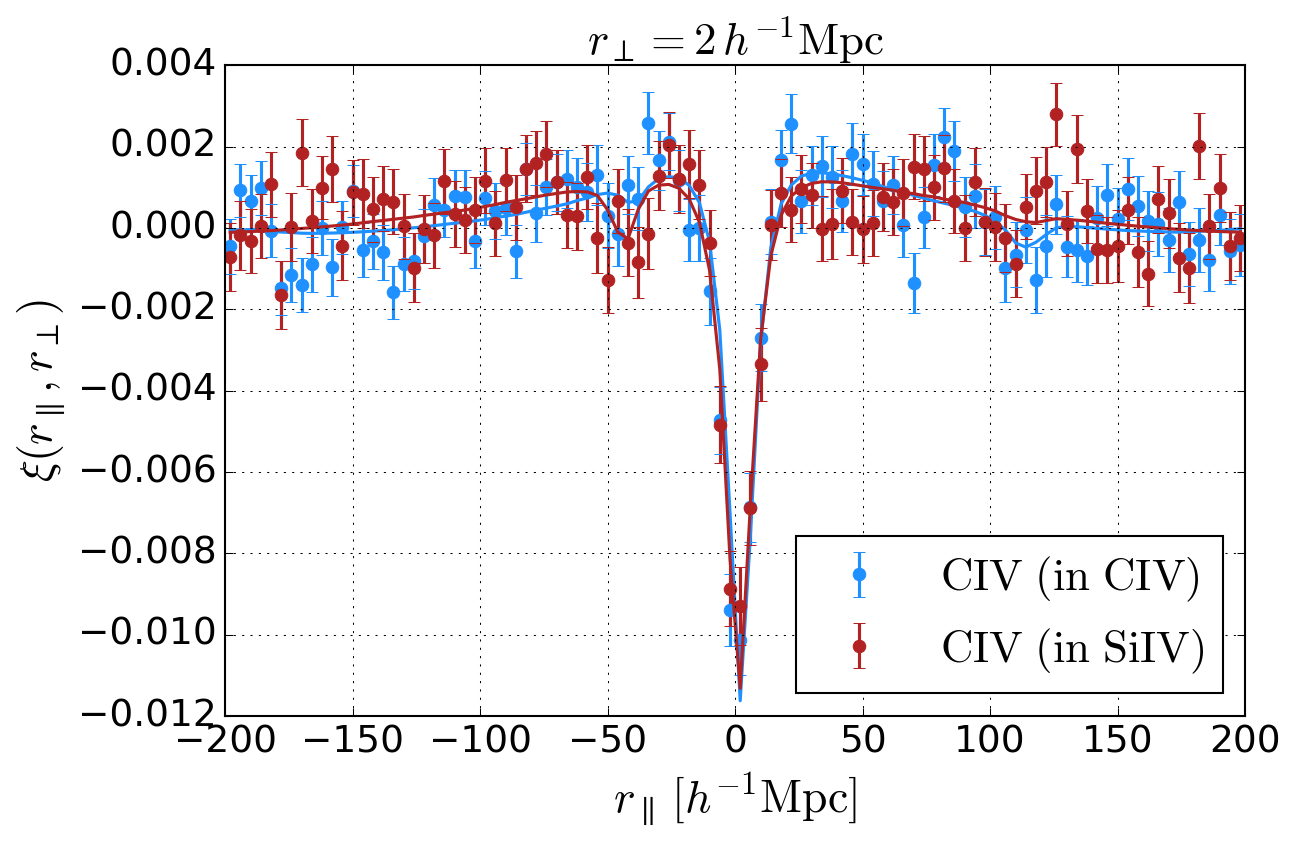}
   \includegraphics[width=3.0in]{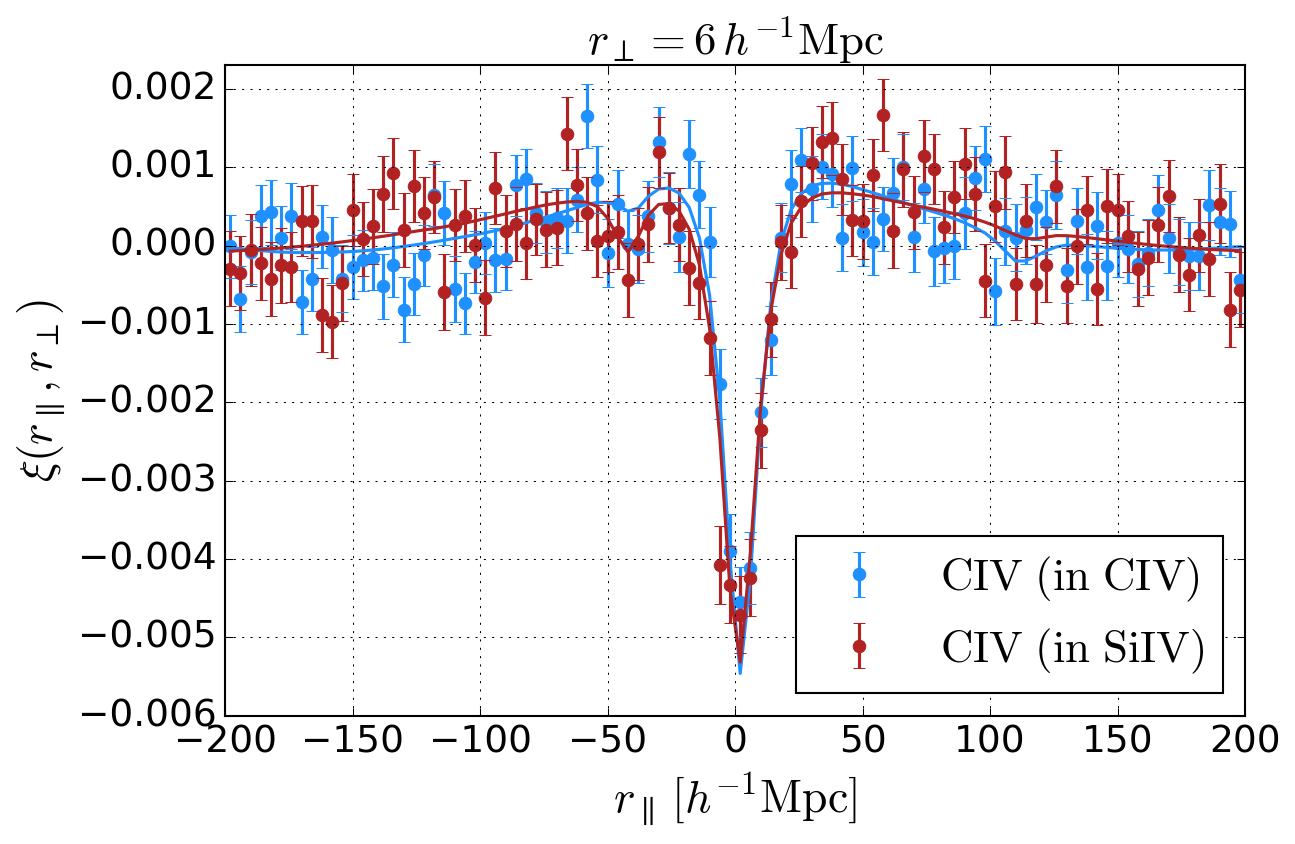}
   \caption{Cross-correlation functions in slices of constant $r_{\perp}$. The blue data points show CIV (in CIV) and the red data points CIV (in SiIV). The solid lines represent the best-fit models obtained for the fitting range $10<r<180~\hMpc$ and extrapolated in the plot to smaller and larger separations. These slices closest to the line of sight ($r_{\perp}=2~\hMpc$ and $r_{\perp}=6~\hMpc$) exhibit correlation features from the metal lines SiII $1526.71~\angstrom$ and FeII $1608.45~\angstrom$. The scale of the correlation function axis is different in the two panels.}
   \label{fig:civslice}
   \end{center}
\end{figure}

\begin{figure}
   \begin{center}
   \includegraphics[width=3.0in]{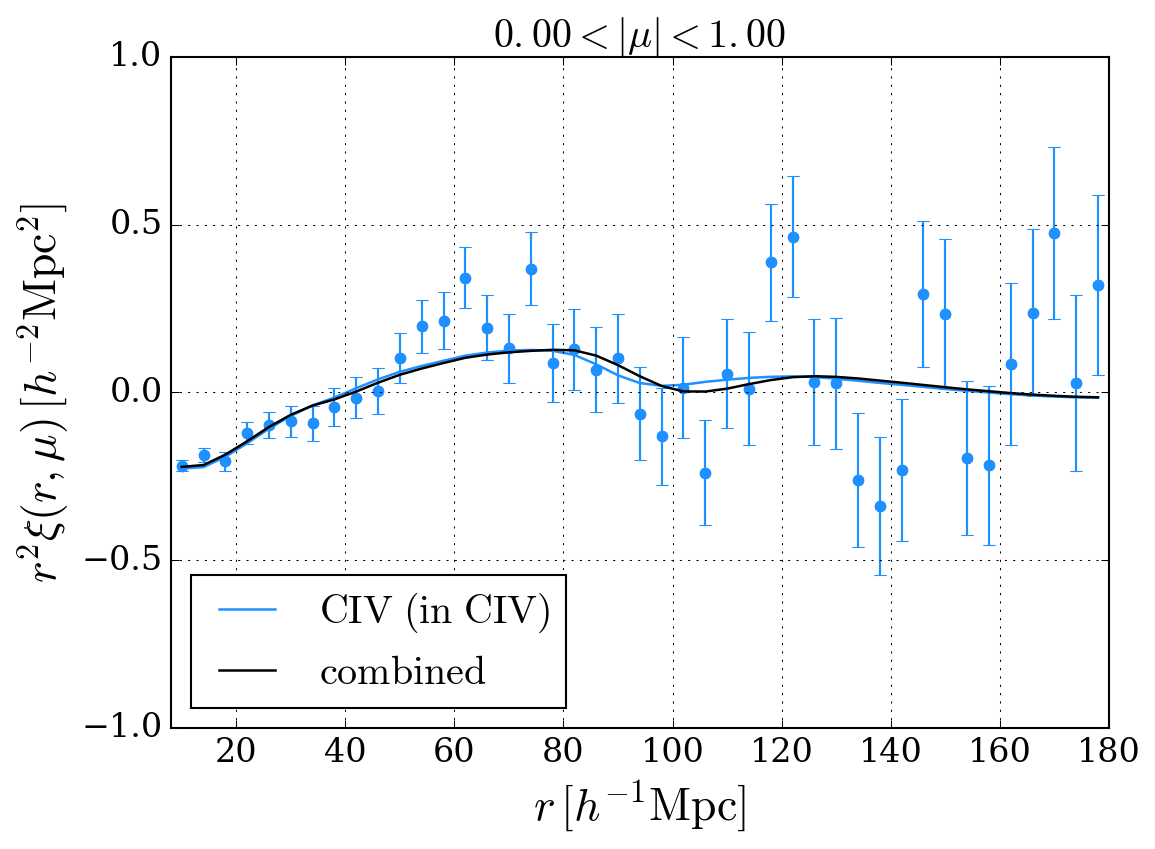}
   \includegraphics[width=3.0in]{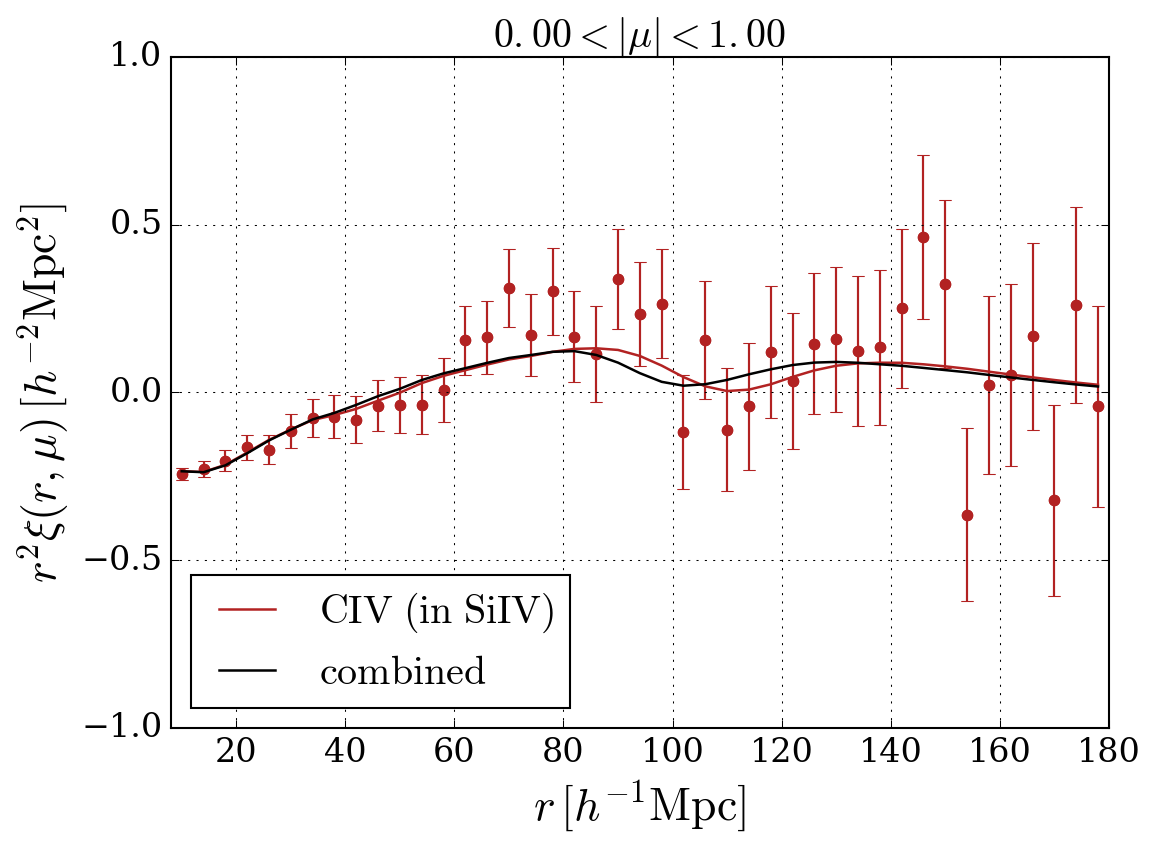}
   \caption{The $r^{2}$-weighted cross-correlation functions for the fitting range $10<r<180~\hMpc$ averaged over the full angular range $0<|\mu|<1$. The data and the best-fit model for CIV (in CIV) are shown in the left panel and for CIV (in SiIV) in the right panel. The black lines indicate the best-fit models for the combined fit.}
   \label{fig:civwedge}
   \end{center}
\end{figure}

Fits to the cross-correlation function are performed using the publicly available code \texttt{picca}\footnote{\url{https://github.com/igmhub/picca/}}. We use the full angular range $-1<\mu<1$, but limit the separations to $10<r<180~\hMpc$, which results in 3180 data bins included in each fit. The model described in section~\ref{sec:model} is fit to the measured cross-correlations for CIV absorption in the CIV forest, the SiIV forest, and the \Lya forest. Table~\ref{table:bestfits} presents the best-fit results. Parameter errors are purely statistical and correspond to $\Delta\chi^{2}=1$. Redshift-dependent parameters are quoted at the effective redshift of each measurement. Figure~\ref{fig:civsiivlyawedge} shows the cross-correlation functions on small scales ($r<40~\hMpc$) measured in the three different forests, averaged over all directions. Correlation signal from CIV absorption is detected in all forests, with the best-fit models (extrapolated to $r<10~\hMpc$ in the figure) being statistically consistent. The best fits to the cross-correlations for CIV absorption in the CIV forest and the SiIV forest are presented in figure~\ref{fig:civslice} for the two lowest $r_{\perp}$ bins and in figure~\ref{fig:civwedge} averaged over all directions. A combined fit to these two cross-correlations is also performed to improve the parameter constraints. The measurement of the cross-correlation for CIV absorption in the \Lya forest is considerably noisier and not included in the combined fit. The fit results are described in more detail below.

\begin{table}
\begin{center}
\begin{tabular}{lrrr|r}
\hline
\hline
\noalign{\smallskip}
Parameter & CIV (in CIV) & CIV (in SiIV) & combined & CIV (in Ly$\alpha$) \\
\noalign{\smallskip}
\hline
\noalign{\smallskip}
$\beta_{\rm CIV}$ & $0.35 \pm 0.21$ & $0.21 \pm 0.21$ & $0.27 \pm 0.14$ & $6.12 \pm 14.27$ \\
$b_{\rm CIV}(1+\beta_{\rm CIV})$ & $-0.019 \pm 0.002$ & $-0.018 \pm 0.002$ & $-0.018 \pm 0.001$ & $-0.015 \pm 0.007$ \\
$\alpha$ & $1.077 \pm 0.090$ & $0.940 \pm 0.078$ & $1.017 \pm 0.061$ & $0.872 \pm 0.141$ \\
$\sigma_{v}~[\hMpc]$ & $4.26 \pm 1.35$ & $4.63 \pm 1.52$ & $4.46 \pm 1.00$ & $10.30 \pm 8.41$ \\
$\Delta r_{\parallel,\rm q}~[\hMpc]$ & $-2.20 \pm 0.47$ & $-1.52 \pm 0.56$ & $-1.87 \pm 0.36$ & $1.78 \pm 2.24$ \\
$10^{3}~b_{\rm SiII(1527)}$ & $-0.6 \pm 0.5$ & $-1.5 \pm 0.5$ & $-1.0 \pm 0.3$ & $-1.1 \pm 1.2$ \\
$10^{3}~b_{\rm FeII(1608)}$ & $-0.9 \pm 0.5$ & $-0.3 \pm 0.5$ & $-0.6 \pm 0.4$ & $0.4 \pm 1.2$ \\
\noalign{\smallskip}
\hline
\noalign{\smallskip}
$z_{\rm eff}$ & $2.06$ & $1.93$ & $2.00$ & $1.64$ \\
$\chi_{\rm min}^{2}$ & $3150.47$ & $3049.30$ & $6205.59$ & $3189.55$ \\
$N_{\rm bin}$ & $3180$ & $3180$ & $6360$ & $3180$ \\
$N_{\rm param}$ & $7$ & $7$ & $7$ & $7$ \\
probability & $0.61$ & $0.94$ & $0.91$ & $0.41$ \\
\noalign{\smallskip}
\hline
\end{tabular}
\end{center}
\caption{Best-fit results for CIV (in CIV), CIV (in SiIV) and their combined fit, as well as for CIV (in Ly$\alpha$). Fits are performed to data in the range $10<r<180~\hMpc$. Parameter errors correspond to $\Delta\chi^2=1$.}
\label{table:bestfits}
\end{table}

\subsection{CIV bias parameters}
\label{subsec:biasparam}

The model provides good fits to the cross-correlation for all three forests. Results for CIV absorption in the CIV forest and the SiIV forest are in excellent agreement, but many of the parameters are poorly constrained in these individual fits. The combined fit improves the precision considerably. The marginalized 68.27\% ($\Delta\chi^{2}=1$) and 95.45\% ($\Delta\chi^{2}=4$) constraints on the parameters $b_{\rm CIV}(1+\beta_{\rm CIV})$ and $\beta_{\rm CIV}$ from the combined fit are
\begin{equation}
b_{\rm CIV}(1+\beta_{\rm CIV})(z_{\rm eff}=2.00) = -0.0183\ _{-0.0014}^{+0.0013}\ _{-0.0029}^{+0.0025}\ ,
\end{equation}
\begin{equation}
\beta_{\rm CIV}(z_{\rm eff}=2.00) = 0.27\ _{-0.14}^{+0.16}\ _{-0.26}^{+0.34}\ .
\end{equation}
Redshift-space distortions of the CIV absorption are only moderately manifested in the current measurement. The best-fit $\beta_{\rm CIV}$ is near the values obtained for quasars $\beta_{\rm q}\approx 0.38$ at $z=1.55$ \cite{2017JCAP...07..017L} and $\beta_{\rm q}\approx 0.25$ at $z=2.39$ \cite{2016JCAP...11..060L}, and damped \Lya absorbers (DLAs) $\beta_{\rm DLA}\approx 0.49$ at $z=2.3$ \cite{2018MNRAS.473.3019P}. From the best-fit values of $\beta_{\rm CIV}$ and $b_{\rm CIV}(1+\beta_{\rm CIV})$, we use their correlation coefficent $\rho=-0.808$ to derive the constraint
\begin{equation}
b_{\rm CIV}(z_{\rm eff}=2.00) = -0.0144 \pm 0.0010\ .
\end{equation}
The CIV absorption linear bias is thus a factor $\sim6$ smaller than the \Lya absorption linear bias translated to the same redshift \cite{2017A&A...608A.130D}.

The cross-correlation measurement for CIV absorption in the \Lya forest is noisier, partially because of the smaller sample of quasars ($z_{\rm q}>2.05$), but also because of the large amount of noise induced by the \Lya absorption. As a consequence, the fit parameters are poorly constrained. Widening the fitting range to $5<r<180~\hMpc$ to include smaller separations adds six data bins where the signal-to-noise ratio of the cross-correlation measurement is high, at the expense of potentially introducing non-linear effects near $r\approx5~\hMpc$ that are not properly taken into account in the model. Table~\ref{table:bestfitsrmin5} in section~\ref{sec:append} summarizes the best-fit results for this alternative fitting range. The parameter constraints improve significantly for the cross-correlation of CIV absorption in the \Lya forest.

\begin{figure}
   \begin{center}
   \includegraphics[width=4.0in]{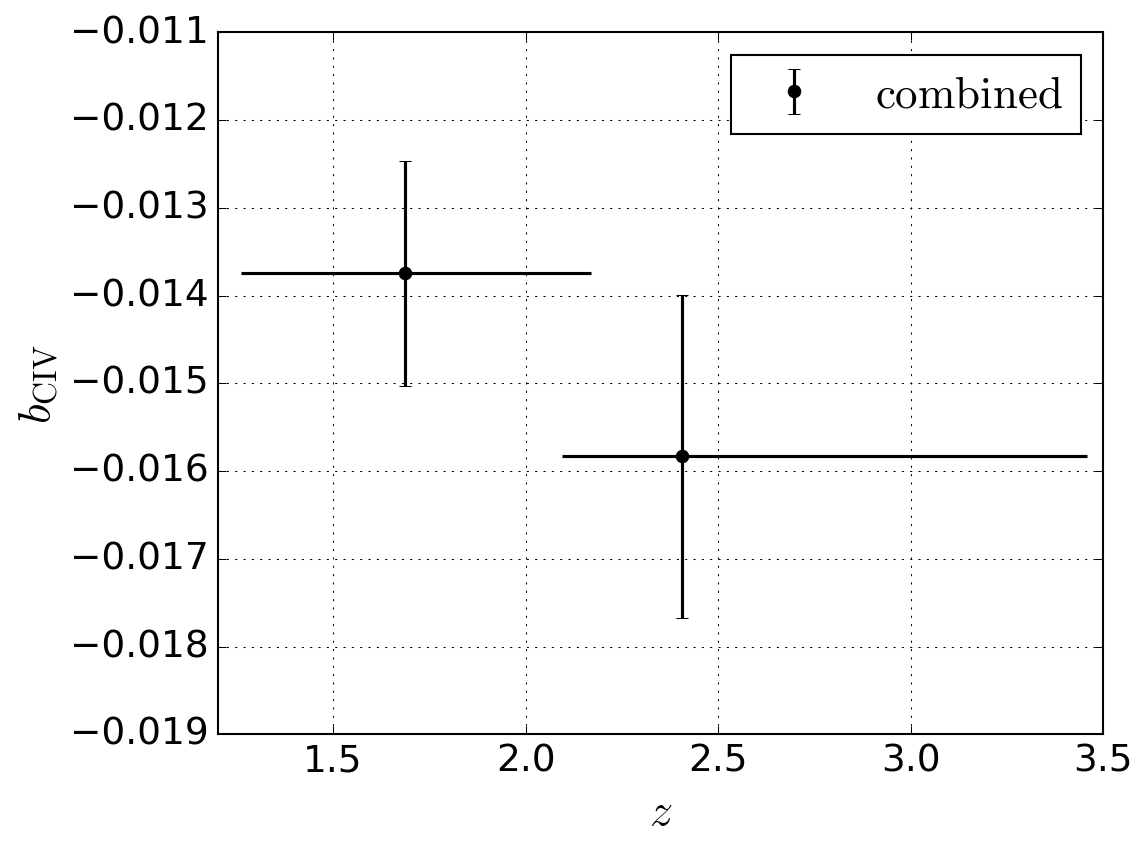}
   \caption{Measurement of the CIV bias parameter from the combined fit to CIV (in CIV) and CIV (in SiIV) for two bins in redshift defined by the data split at $z=2.2$. The horizontal lines indicate the redshift range of each bin.}
   \label{fig:civbias}
   \end{center}
\end{figure}

To search for possible redshift evolution of the CIV absorption linear bias, we split the forest and quasar samples at $z=2.2$. Table~\ref{table:bestfitssplit} presents the results of performing the combined fit to the cross-correlations for each redshift bin. The derived constraints on $b_{\rm CIV}$ at the effective redshifts for the samples with $z<2.2$ ($\rho=-0.857$) and $z>2.2$ ($\rho=-0.753$) are
\begin{equation}
b_{\rm CIV}(z_{\rm eff}=1.69)=-0.0137\pm0.0013
\quad , \quad
b_{\rm CIV}(z_{\rm eff}=2.41)=-0.0158\pm0.0018\ .
\end{equation}
Figure~\ref{fig:civbias} shows the $b_{\rm CIV}$ measurements from the data split. Assuming a power-law evolution as in equation~(\ref{eq:biasevo}), the measurements imply that the exponent $\gamma=0.60\pm0.63$. The statistical error on $\gamma$ is large, but the value differs significantly from the \Lya bias evolution, $\gamma_{\rm Ly\alpha}=2.9$.

\begin{table}
\begin{center}
\begin{tabular}{lccccc}
\hline
\hline
\noalign{\smallskip}
Analysis & $z_{\rm eff}$ & $\beta_{\rm CIV}$ & $b_{\rm CIV}(1+\beta_{\rm CIV})$ & $\chi_{\rm min}^2/DOF$ & prob. \\
\noalign{\smallskip}
\hline
\noalign{\smallskip}
combined $z<2.2$ & $1.69$ & $0.05 \pm 0.19$ & $-0.014 \pm 0.002$ & $6197.27/(6360-7)$ & $0.92$ \\
combined $z>2.2$ & $2.41$ & $0.67 \pm 0.29$ & $-0.026 \pm 0.003$ & $6321.14/(6360-7)$ & $0.61$ \\
\noalign{\smallskip}
\hline
\end{tabular}
\end{center}
\caption{Best-fit results for the two main parameters for the combined fit to CIV (in CIV) and CIV (in SiIV) with a data split at redshift $z=2.2$. Parameter errors correspond to $\Delta\chi^2=1$.}
\label{table:bestfitssplit}
\end{table}

\subsection{Isotropic BAO}
\label{subsec:isobao}

Only a moderate BAO signal is observed in the current cross-correlation measurement, as illustrated in figure~\ref{fig:civwedge}, even though we include quasar spectra over a wide redshift range ($1.4<z<3.5$). The error on the isotropic BAO parameter $\alpha$ given in table~\ref{table:bestfits} is derived using $\Delta\chi^{2}=1$, nominally corresponding to the $68.27\%$ confidence limit. As discussed in \cite{2017A&A...608A.130D}, the correspondence between $\Delta\chi^{2}$ and confidence limits is not expected to be exact because the model is not a linear function of $\alpha$. Following the method in \cite{2017A&A...608A.130D} for estimating the effective mapping between $\Delta\chi^{2}$ and confidence limits, we generate 500 simulated cross-correlation functions by adding noise, drawn randomly from the measured covariance matrix, to the fiducial cosmological model with the best-fit values of non-BAO parameters from the combined fit. From the resulting distribution, we derive that $\Delta\chi^{2}=(1.19,4.17)$ corresponds to the standard $1\sigma$ and $2\sigma$ confidence limits of ($68.27\%, 95.45\%$).

\begin{figure}
   \begin{center}
   \includegraphics[width=4.0in]{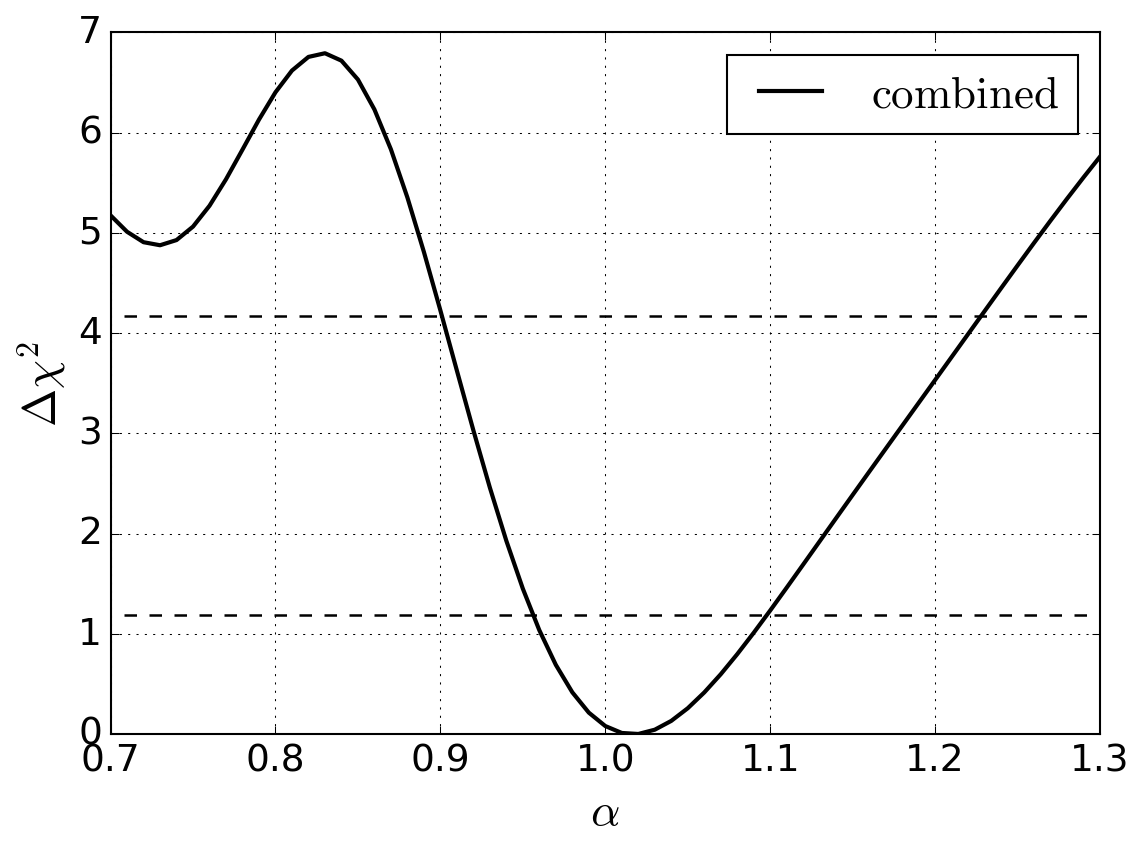}
   \caption{Constraint on the isotropic BAO parameter from the combined fit to CIV (in CIV) and CIV (in SiIV). The dashed horizontal lines indicate $\Delta\chi^{2}=1.19$ and $\Delta\chi^{2}=4.17$ corresponding to $68.27\%$ and $95.45\%$ C.L.. The secondary minumum near $\alpha=0.73$ corresponds to $r\sim140~\hMpc$, far from the expected peak position.}
   \label{fig:civaiso}
   \end{center}
\end{figure}

Figure~\ref{fig:civaiso} shows the constraint on the isotropic BAO parameter for the combined fit. Some asymmetry is noticable in the curve. The secondary minimum near $\alpha=0.73$ corresponds to $r\sim140~\hMpc$, far from the expected peak position. The best-fit value and ($68.27\%, 95.45\%$) confidence limits are
\begin{equation}
\alpha(z_{\rm eff}=2.00)=1.017\ _{-0.061}^{+0.081}\ _{-0.116}^{+0.211}\ ,
\label{eqn:bestfitbao}
\end{equation}
consistent with the fiducial BAO peak position. We quantify the significance of the BAO peak by fitting a no-peak model ($A_{\rm peak}=0$) for the combined fit; this fit yields $\Delta\chi_{\rm min}^2=3.22$ compared to the standard fit with the BAO peak included, corresponding to a detection significance of $1.7\sigma$.

\subsection{Future constraints on isotropic BAO}
\label{subsec:baoproject}

The results in section~\ref{subsec:isobao} demonstrate that the quasar-CIV cross-correlation has the potential to robustly probe BAO in next-generation quasar absorption surveys. Here, we aim to estimate the expected precision of the BAO peak position measurement that can be achieved with the data that will be available in the final samples from eBOSS and DESI. These projections follow simple scaling arguments based on the planned area and quasar number density of each survey. We only consider constraints obtained from the cross-correlation, since the auto-correlation of the CIV transmission field yields no detectable correlation signal using the current data set.

The covariance scales as the inverse of the number of quasar-pixel pairs, thus we expect a scaling of the covariance that is linear with the survey area, tracer quasar number density, and forest quasar number density. Because the scaling cancels any multiplicative correction factor on the number density (such as removing BALs that account for about 10\% of the sample), we set the tracer quasar and forest quasar number densities to be identical. The decreasingly significant sample of SDSS DR7 quasars, which are only used as tracer quasars, is neglected. The scaling of the covariance matrix is thus assumed to be linear in survey area and quadratic in quasar number density.

Whereas the main results presented in this work are derived using quasars over the wide redshift range $1.4<z<3.5$ to reduce the statistical uncertainties, future measurements of the cross-correlation and its BAO peak will focus on the narrower range $1.4<z<2.2$ as the sample of quasars in this range rapidly increases, because of the eBOSS target selection procedure designed to favor measurements of the quasar auto-correlation. As the basis for our projections, we therefore consider the cross-correlation measurements for CIV absorption in the CIV forest and the SiIV forest for $1.4<z<2.2$ that were used in the data split in section~\ref{subsec:biasparam}. We assume that these cross-correlations are primarily determined by the current eBOSS quasars. This assumption is reasonable, because the area that includes eBOSS quasars is a few times larger than that including SEQUELS quasars, and the number density in the eBOSS area is a few times higher than the number density in the area covered solely by BOSS+DR7 quasars. The measured covariance matrices are scaled by the appropriate factors, $s$, to simulate the final samples from eBOSS and DESI. From the scaled covariance matrices and the fiducial cross-correlation model of the combined fit, we generate 500 simulated cross-correlation functions using the same method as in section~\ref{subsec:isobao}.

\begin{figure}
   \begin{center}
   \includegraphics[width=4.0in]{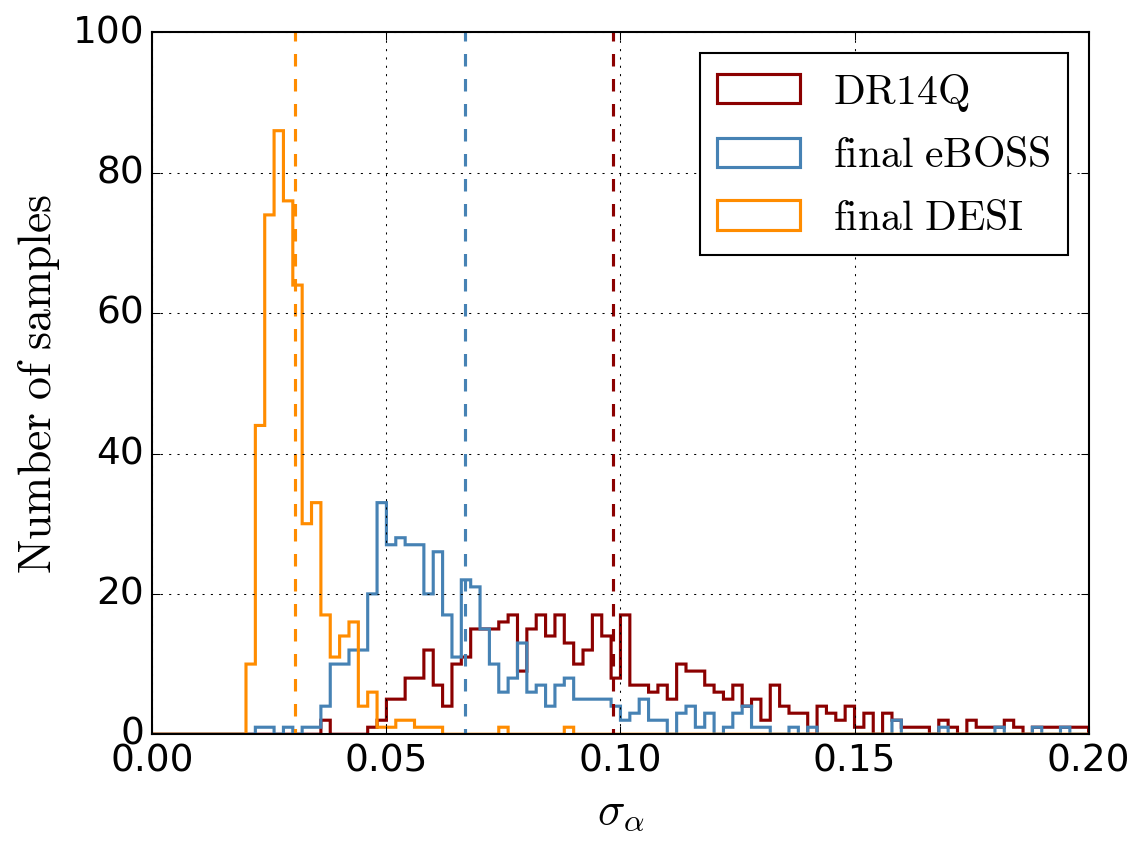}
   \caption{Distribution of the error on the isotropic BAO parameter obtained from the combined fit to CIV (in CIV) and CIV (in SiIV) for 500 realisations with $1.4<z<2.2$. The histogram for DR14Q is shown in red, for final eBOSS in blue, and for final DESI in orange. The dashed horizontal lines indicate the mean of the distributions: $\langle\sigma_{\alpha}\rangle=0.098$ (DR14Q), $0.067$ (final eBOSS), $0.031$ (final DESI).}
   \label{fig:aisoprojection}
   \end{center}
\end{figure}

Figure~\ref{fig:aisoprojection} presents the distribution of the error on the isotropic BAO parameter obtained from the combined fits for  the 500 realisations using the DR14Q covariance matrix. The distribution is wide, with a mean value of $\langle\sigma_{\alpha}\rangle=0.098$. This value is in good agreement with the constraint obtained for the combined fit for $1.4<z<2.2$ in table~\ref{table:bestfitssplit},
\begin{equation}
\alpha(z_{\rm eff}=1.69)=0.997_{\ -0.067}^{\ +0.107}\ .
\label{eqn:bestfitbaolowz}
\end{equation}

The eBOSS quasar clustering program is designed to target and measure redshifts for approximately 500,000 quasars in the redshift range $0.9<z<2.2$ over a survey area of $7500~{\rm deg}^2$. Scaling from the current sample in DR14Q to the final eBOSS sample simply involves taking the ratio of the areas, which can be approximated with the ratio of the sample sizes. There are 111,601 eBOSS quasars in the range $0.9<z<2.2$ in DR14Q, making the scale factor $s = 0.22$. The $\sigma_{\alpha}$ distribution obtained from the combined fits to the simulated cross-correlation functions is displayed in figure~\ref{fig:aisoprojection}. The expected final eBOSS BAO error is $\langle\sigma_{\alpha}\rangle=0.067$.

In order to scale to the final DESI sample, we adopt the DESI target numbers of measuring quasar spectra over a survey area of $14,000~{\rm deg}^2$ with a number density of $120~{\rm deg}^{-2}$ for $0.9<z<2.1$. There are 106,045 eBOSS quasars in the corresponding redshift range in DR14Q. Scaling linearly in survey area and quadratically in quasar number density yields the factor $s = 0.033$. The final DESI sample can achieve a precision of 
$\langle\sigma_{\alpha}\rangle=0.031$.

\section{Discussion}
\label{sec:discussion}

This paper explores the large-scale structure tracer potential of triply-ionized carbon (CIV) absorption along the line of sight to background quasars using SDSS-IV/eBOSS DR14, an idea initially presented in \cite{2014MNRAS.445L.104P}. We demonstrate that such a measurement of baryon acoustic oscillations is possible. This is a non-trivial matter since the carbon that gives rise to this absorption is present due to astrophysical processes such as metal production in stars, supernova ejecta both dispersing these metals and driving them from galaxies, the mass loading that couples this energy into an entrained flow out of the source galaxy, the halo potential well that defines whether the metals have become unbound, and finally the velocity sheer sensitive to the competition between outflow expansion and gas return. Furthermore, the ionization properties that define the incidence of the CIV species are potentially sensitive to UV background fluctuations due to proximity to galaxies and quasars in a manner distinct from that of Ly$\alpha$. The physical model (described in section~\ref{sec:model}) used to describe this tracer is an excellent fit to the CIV forest data, despite the fact that these physical processes are neglected and that the model is only a small modification of that used for \Lya forest studies. This supports our hypothesis that CIV absorption is indeed a viable tracer of large-scale structure.

We measure the cross-correlation with quasars, taking advantage of both the relatively high bias of quasars (as a boost to the correlation signal), and the fact that some systematic errors in an auto-correlation do not produce bias in cross-correlation (e.g. galaxy sample impurity, galaxy selection function and pixel-level spectroscopic correlations other than those tracing the intended large-scale structures). This quasar-CIV cross-correlation is computed using CIV absorption in three distinct bands: the `in CIV' band (named because it is dominated by CIV absorption), the `in SiIV' band (as there is also significant absorption by triply ionized silicon), and the `in Ly$\alpha$' band (which includes absorption from both \Lya and SiIV). The probabilities of fits to all three measurements are individually good, again, indicating that our physical model performs well. We find that the best-fitting models for the CIV (in CIV) and CIV (in SiIV) are in good agreement and furthermore we perform a combined fit which also shows a good quality of fit. The measurement using the CIV (in Ly$\alpha$) band is noisier than that of the other two bands such that the values of $b_{\rm CIV}$ and $\beta_{\rm CIV}$ are largely unconstrained. Only the optimal combination of $b_{\rm CIV}(1+\beta_{\rm CIV})$ is constrained and this value, while noisy, is consistent with the measurements from CIV (in CIV) and CIV (in SiIV). Given this lack of constraining power in current data, we do not use the CIV (in Ly$\alpha$) measurement for further results in the paper and focus instead on the combination of  CIV (in CIV) and CIV (in SiIV).

\subsection{BAO measurements}
\label{subsec:baodisc}

We estimate the BAO scale, as quantified by the isotropic scale parameter, as $\alpha = 1.017 \pm 0.061$ using the entire redshift range available to us of $1.4<z<3.5$ with an effective redshift $z_{\rm eff}=2.00$, where the error is derived by the fitter alone. We improve on this by both exploring the appropriate $\Delta \chi^2$ corresponding to $1\sigma$ and $2\sigma$ confidence limits and scanning the $\chi^2$ surface. As a result, we derive a more refined and slightly larger estimate of uncertainties as given in equation~(\ref{eqn:bestfitbao}). We do not pursue a detailed comparison with mock data or a subsequent exploration of derived cosmological constraints for various reasons. This is chiefly because the BAO detection associated with the constraint above fails to reach the $2\sigma$ level based on current data.

In the current work, we have attempted to measure a single BAO across our redshift bin. Since the redshift range is wide, the evolution of the BAO scale (in physical units, i.e. degrees and redshift differences) will vary significantly across the redshift window. If the true cosmological model is not consistent with the fiducial model used to assign distances to redshifts and angles, this could lead to smoothing and hence degradation in the BAO signal-to-noise ratio. A more sophisticated treatment of redshift evolution will be warranted as the data quality improves.

The redshift interval $1.4<z<3.5$ overlaps significantly with other analyses. There is a weak redshift overlap with the mature and more constraining analysis of \Lya auto-correlations \cite{2017A&A...603A..12B} and the \Lya cross-correlation with quasars \cite{2017A&A...608A.130D} in BOSS DR12 data, with effective redshift of $z\approx 2.4$ and minimum redshifts of $z\approx 2$. If stated as an isotropic constraint of the BAO scale, these \Lya measurements combined would reach $\approx 2\%$ precision. The above redshift range also overlaps with the eBOSS DR14 quasar auto-correlation measurement (for which the low-$z$ sample was constructed) \cite{2017arXiv170506373A}, which is performed for quasars at $0.8<z<2.2$ and produces a 3.8\% precision measurement of isotropic BAO. The long-term goal is to complement this lower redshift measurement, so we re-assess our BAO constraint using the same maximum quasar redshift. This choice effectively sets a cosmological analysis range of $1.4<z\lesssim2.1$ for the potentially constraining quasar-CIV cross-correlation and the resultant BAO precision is approximately 10\% here. This weak constraint on current data further reinforces our decision to forego detailed BAO interpretation, but these measurements serve as a demonstration.

If the current uncertainties for both measurements scale as the sample grows and if the errors between the quasar auto- and the CIV cross-correlation are orthogonal, then the CIV cross-correlation can provide a non-negligible boost to the BAO precision (the current measurement of 3.8\% precision would be improved to $\approx 3.6\%$ assuming uncorrelated errors). More importantly, a comparison would provide a consistency check between BAO as measured by tracer galaxies (in the broadest sense) and BAO as measured in quasar absorption. To date, quasar absorption BAO has been limited to $z>2$ and galaxy BAO has been limited to $z<2$. The measurement using CIV as a large-scale structure tracer bridges this redshift divide, for the first time allowing a consistency check between these methodologically different approaches to measuring BAO.

In order to assess the potential constraining power of the CIV absorption cross-correlated with quasars at $z<2.2$, we perform a projection both for the final projected eBOSS and the final DESI survey (section~\ref{subsec:baoproject}) by generating random realisations of the correlation function from a scaled version of our covariance matrix and applying our correlation function fitter. The scaling is based on simple scaling relations with respect to survey area and density. We conclude that the projected mean CIV BAO precision for eBOSS is $\sim7\%$ and that for DESI is $\sim3\%$, based on the mean precision of 500 random survey realisations.  Mock-to-mock variations are significant in figure~\ref{fig:aisoprojection}, meaning that the precision of a real-world measurement may differ substantially from this. In light of the fact that approximately a quarter of eBOSS data is in hand, the degree of uncertainty in the BAO precision of final eBOSS is less free than this analysis indicates.

These projections refer only to the cross-correlation of CIV forest absorption for `in CIV' and `in SiIV' wavelength ranges, but as the signal-to-noise ratio improves other absorption wavelength ranges will provide useful constraints such as CIV in Ly$\alpha$ and CIV in the Ly$\beta$ forests. Moreover, other species will become constraining and add value to these $z<2$ measurements such as SiIV (as a tracer and not a source of noise as in this work) and MgII. Finally, metal auto-correlation measurements may also add value. As such, these projected error estimates for  $z<2$ absorption-based BAO measurements should be viewed as conservative.

\subsection{Other parameter constraints}
\label{subsec:astrodisc}

A number of parameters are measured in our analysis in addition to the isotropic BAO scale parameter. These parameters are interesting in their own right, such as the linear density bias of the CIV forest ($b_{\rm CIV}$), its redshift evolution ($\gamma$), the redshift-space distortions of the CIV forest ($\beta_{\rm CIV}$), and the systematic and stochastic CIV line-of-sight shift with respect to measured quasar redshift ($\Delta r_{\parallel, \rm q}$ and $\sigma_{v}$, respectively).

The linear density bias as applied for absorption forests differs from the galaxy bias since linear bias of an absorption forest as a whole describes the full range of density fluctuations for the entire line-of-sight skewer of gas. This linear bias is a combination of the forest opacity and the bias of the gas clouds giving rise to this absorption. A substantial literature exists focussed on exploring the meaning of biases in the \Lya forest \cite{2003ApJ...585...34M, 2011JCAP...09..001S, 2012JCAP...03..004S,2016JCAP...03..016C}, which also largely apply to the CIV forest with different values describing the distinct way that CIV traces structure. An additional weak bias must also be present that describes the inhomogeneity of metal enrichment, but we neglect this contribution as a small scale effect here (a choice which appears justified by quality of our fits).

Our measured CIV linear bias is approximately an order of magnitude smaller than the measured value yielded by the \Lya forest - a result to be expected in light of prior results \cite{2014MNRAS.445L.104P}. A potentially more enlightening measurement, however, is its apparent redshift evolution as shown in figure~\ref{fig:civbias} and reflected in the evolution between the high and low redshift fits, $\gamma=0.60\pm0.63$. This behaviour is quite different from the value $\gamma_{\rm Ly\alpha}=2.9$ of the \Lya forest, albeit measured at somewhat higher redshifts. In light of the well-documented lack of strong evolution in the cosmic mass density of CIV $\Omega_{\rm CIV}$ at $z\sim2$ \cite{2001ApJ...561L.153S, 2010MNRAS.401.2715D, 2013ApJ...763...37C}, this alternative measure of evolution in the CIV population warrants further investigation. It should be noted that both the CIV forest and the \Lya forest mix together different types of gas conditions (e.g. inflow/outflow, optically thick/thin, collisionally ionized or photoionized and the presence of damping wings) with subtle differences in weighting for both the linear bias and the cosmic mass density of a species. Furthermore, evolution in both these quantities may not be a simple consequence of changing enrichment but also changing ionization conditions.

\begin{figure}
   \begin{center}
   \includegraphics[width=4.0in]{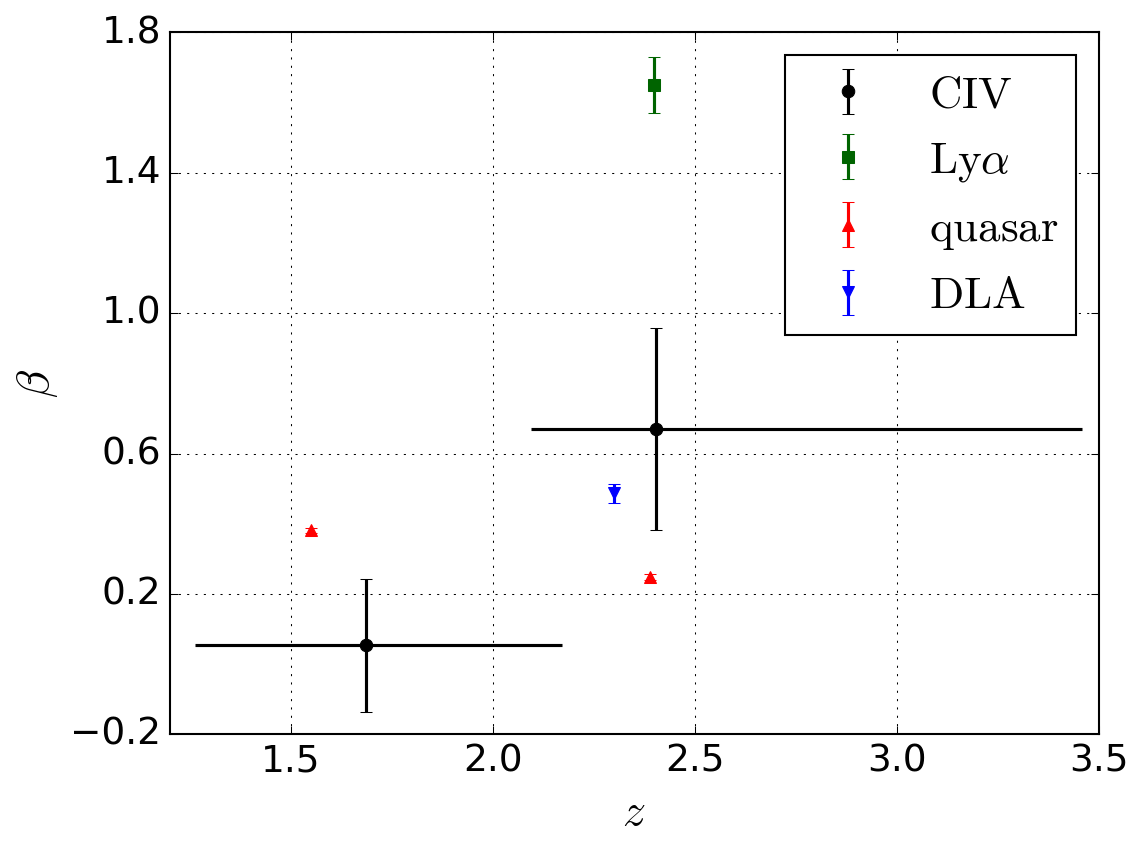}
   \caption{Measurements of the redshift-space distortion parameter of different large-scale structure tracers. The black points show the $\beta_{\rm CIV}$ measurements from the combined fit to CIV (in CIV) and CIV (in SiIV) for two bins in redshift defined by the data split at $z=2.2$ (this work). The horizontal lines indicate the redshift range of each bin. Other measurements come from the \Lya forest \cite{2017A&A...608A.130D}, quasars \cite{2016JCAP...11..060L,2017JCAP...07..017L} and DLAs \cite{2018MNRAS.473.3019P}.}
   \label{fig:civbeta}
   \end{center}
\end{figure}

Redshift-space distortions are formally independent of the atomic line properties that set the opacity of the forest studied (such as oscillator strength) and so allow us to consider a more generally comparable quantity. We provide a marginal detection of non-zero redshift-space distortions as quantified by the parameter $\beta$; low positive values of $\beta$ are generically indicative of collapsed structures, while high values reflect large-scale collapse. Negative values could in principal occur in a regime where galactic outflows dominate. Figure~\ref{fig:civbeta} presents our redshift-space distortions compared with that of quasars \cite{2016JCAP...11..060L,2017JCAP...07..017L}, DLAs \cite{2018MNRAS.473.3019P} and the \Lya forest \cite{2017A&A...608A.130D}. Despite large uncertainty, our measurements appear to reflect redshift-space distortions that disfavour the picture of large-scale gas collapse of the \Lya forest and are broadly characteristic of collapsed structures such as DLAs and quasars. These results are also suggestive of a measured transition over cosmic time towards greater collapsed structure dominance in the CIV forest. We would stress, however, that all quoted measurements of redshift-space distortions reflect the ensemble average of each population and any relative weighting. In addition, $\beta$ as measured in absorption can also show some dependence on the small-scale smoothing due to intergalactic medium temperature and pressure \cite{2011JCAP...09..001S}. It is also possible that the redshift-space distortion effect of infall on large scales is being counter-balanced to some extent by the opposing process of galactic outflows amongst our ensemble properties. With these caveats in mind, we may combine our measurements of redshift-space distortions and linear bias in order to infer the associated ensemble halo bias $b_{\rm h}=f/\beta_{\rm CIV}$ and the CIV mean effective optical depth $\overline{\tau}=-b_{\rm CIV}\beta_{\rm CIV}/f$ \citep{2012JCAP...07..028F}. From the best-fit values for the $z>2.2$ sample in Table~\ref{table:bestfitssplit}, we derive $b_{\rm h}=1.45\pm0.63$ and $\overline{\tau}=0.011\pm0.004$ at the mean redshift $z=2.41$. The derived halo bias does not differ significantly from the DLA bias \citep{2018MNRAS.473.3019P}, suggesting that the characteristic halo mass which reflects the CIV absorption hosting halo population is less than or similar to that of the DLA hosting halo populations. The mean opacity is broadly consistent with the CIV population measurements \cite{2014MNRAS.445L.104P}. The smaller values of $b_{\rm CIV}$ and $\beta_{\rm CIV}$ obtained for the $z<2.2$ sample suggests evolution with cosmic time towards larger halo bias and smaller mean opacity, but more precise measurements are needed in order to draw firm conclusions. Once more, this is an area that warrants further investigation, particularly in light of growing massive spectroscopic surveys of the intergalactic medium through DESI and WEAVE \cite{2016sf2a.conf..259P}.

A further potential effect influencing our measurement of the large-scale distribution of CIV concerns the question of UV background inhomogeneity and how it impacts upon the fraction of carbon in the triply-ionized state. In our fits, we have the freedom to describe a large-scale radial dependence arising from UV background inhomogeneity \cite{2014MNRAS.442..187G, 2014PhRvD..89h3010P}, which has been applied to the observed cross-correlation of \Lya and quasars \cite{2017A&A...608A.130D}. We allow the bias term associated with UV background inhomogeneity to float in our fits but the best-fit values are consistent with no fluctuations in the current data. UV background inhomogeneity may also present itself as systematic line-of-sight asymmetry in the CIV distribution with respect to quasar redshift due to the impact of light travel time and quasar duty cycles \cite{2008ApJ...674..660V}. This effect is included in our fits as a contribution to $\Delta r_{\parallel,\rm q}$, though this parameter also describes the systematic error in quasar redshifts. In our low-$z$ sample this constitutes a blueshift of $\approx 260$~km~s$^{-1}$. Comparing our measurement with the current understanding of quasar redshift systematics \cite{2016AJ....151...44D, 2017arXiv171205029P}, we find the offset to be large, but a greater understanding of redshift systematics is needed in order to draw conclusions from this measurement. 

Another study measuring the cross-correlation of quasars and the CIV forest has been completed on a very similar timescale \cite{2017arXiv171209886G}. It used a smaller sample, SDSS-III DR12, and focussed on measuring the CIV forest bias (and combining it with measurements of the mean opacity from other data) in order to understand the halos that dominate the CIV signal. Since their data sample is dominated by $z>2.1$ quasars, we can compare our results for the high-$z$ sample to theirs at a similar mean redshift. Their best-fit values $b_{\rm CIV}(1+\beta_{\rm CIV})=-0.024\pm0.003$ and $\beta_{\rm CIV}=1.1\pm0.6$ at the mean redshift $z=2.3$ are in good agreement with our measurements.

\section{Summary}
\label{sec:summary}

We use quasars in the redshift range $1.4<z<3.5$ from SDSS-IV DR14, including a large sample of new quasars from the first two years of eBOSS observations, to measure the cross-correlation of CIV absorption with quasar positions. By performing a combined fit to the cross-correlation measurements for CIV absorption in the CIV forest and the SiIV forest, we measure $\beta_{\rm CIV}=0.27_{\ -0.14\ -0.26}^{\ +0.16\ +0.34}$ and $b_{\rm CIV}(1+\beta_{\rm CIV})=-0.0183_{\ -0.0014\ -0.0029}^{\ +0.0013\ +0.0025}$ ($1\sigma$ and $2\sigma$ statistical errors) at the effective redshift $z_{\rm eff}=2.00$. The data split at $z=2.2$ indicates a weak redshift evolution of the CIV linear bias as described by the power-law exponent $\gamma=0.60\pm0.63$. These measurements of bias and redshift-space distortions hint at valuable alternative measurements of the CIV population (and its comparison to the underlying density distribution) as the eBOSS sample grows.

The fit for the isotropic BAO scale yields only a moderate peak significance ($1.7\sigma$) for the data available in DR14, but serves as a demonstration that the quasar-CIV cross-correlation has the potential to robustly probe BAO in next-generation quasar absorption surveys. By scaling the covariance matrix of the current measurement to simulate the expected future cross-correlation measurements, we estimate that the final quasar samples for redshifts $1.4<z<2.2$ from eBOSS and DESI can measure the isotropic BAO scale to $\sim7\%$ and $\sim3\%$ precision, respectively, at $z\simeq1.6$.

\section*{Acknowledgments}

The authors would like to thank Andreu Font-Ribera and Christophe Y{\`e}che for their useful comments.

This work was supported by the A*MIDEX project (ANR-11-IDEX-0001-02) funded by the ``Investissements d'Avenir'' French Government program, managed by the French National Research Agency (ANR), and by ANR under contract ANR-14-ACHN-0021.

Funding for the Sloan Digital Sky Survey IV has been provided by the Alfred P. Sloan Foundation, the U.S. Department of Energy Office of Science, and the Participating Institutions. SDSS acknowledges
support and resources from the Center for High-Performance Computing at the University of Utah. The SDSS web site is \url{http://www.sdss.org/}.

SDSS is managed by the Astrophysical Research Consortium for the Participating Institutions of the SDSS Collaboration including the Brazilian Participation Group, the Carnegie Institution for Science, Carnegie Mellon University, the Chilean Participation Group, the French Participation Group, Harvard-Smithsonian Center for Astrophysics, Instituto de Astrof{\'i}sica de Canarias, The Johns Hopkins University, Kavli Institute for the Physics and Mathematics of the Universe (IPMU) / University of Tokyo, Lawrence Berkeley National Laboratory, Leibniz Institut f{\"u}r Astrophysik Potsdam (AIP), Max-Planck-Institut f{\"u}r Astronomie (MPIA Heidelberg), Max-Planck-Institut f{\"u}r Astrophysik (MPA Garching), Max-Planck-Institut f{\"u}r Extraterrestrische Physik (MPE), National Astronomical Observatories of China, New Mexico State University, New York University, University of Notre Dame, Observat{\'o}rio Nacional / MCTI, The Ohio State University, Pennsylvania State University, Shanghai Astronomical Observatory, United Kingdom Participation Group, Universidad Nacional Aut{\'o}noma de M{\'e}xico, University of Arizona, University of Colorado Boulder, University of Oxford, University of Portsmouth, University of Utah, University of Virginia, University of Washington, University of Wisconsin, Vanderbilt University, and Yale University.


\appendix
\section{Results for alternative fitting range}
\label{sec:append}

\begin{table}[htb]
\begin{center}
\begin{tabular}{lrrr|r}
\hline
\hline
\noalign{\smallskip}
Parameter & CIV (in CIV) & CIV (in SiIV) & combined & CIV (in Ly$\alpha$) \\
\noalign{\smallskip}
\hline
\noalign{\smallskip}
$\beta_{\rm CIV}$ & $0.39 \pm 0.21$ & $0.15 \pm 0.18$ & $0.26 \pm 0.14$ & $0.50 \pm 0.82$ \\
$b_{\rm CIV}(1+\beta_{\rm CIV})$ & $-0.020 \pm 0.002$ & $-0.018 \pm 0.002$ & $-0.019 \pm 0.001$ & $-0.013 \pm 0.004$ \\
$\alpha$ & $1.074 \pm 0.088$ & $0.936 \pm 0.077$ & $1.017 \pm 0.061$ & $0.869 \pm 0.103$ \\
$\sigma_{v}~[\hMpc]$ & $5.70 \pm 1.22$ & $5.68 \pm 1.29$ & $5.73 \pm 0.89$ & $2.90 \pm 3.00$ \\
$\Delta r_{\parallel,\rm q}~[\hMpc]$ & $-1.91 \pm 0.36$ & $-1.08 \pm 0.41$ & $-1.53 \pm 0.27$ & $-1.16 \pm 1.22$ \\
$10^{3}~b_{\rm SiII(1527)}$ & $-0.6 \pm 0.5$ & $-1.5 \pm 0.5$ & $-1.1 \pm 0.3$ & $-0.6 \pm 1.2$ \\
$10^{3}~b_{\rm FeII(1608)}$ & $-0.9 \pm 0.5$ & $-0.3 \pm 0.5$ & $-0.6 \pm 0.4$ & $0.3 \pm 1.2$ \\
\noalign{\smallskip}
\hline
\noalign{\smallskip}
$z_{\rm eff}$ & $2.06$ & $1.92$ & $1.99$ & $1.64$ \\
$\chi_{\rm min}^{2}$ & $3157.55$ & $3066.32$ & $6231.03$ & $3202.96$ \\
$N_{\rm bin}$ & $3186$ & $3186$ & $6372$ & $3186$ \\
$N_{\rm param}$ & $7$ & $7$ & $7$ & $7$ \\
probability & $0.60$ & $0.92$ & $0.88$ & $0.38$ \\
\noalign{\smallskip}
\hline
\end{tabular}
\end{center}
\caption{Same as table~\ref{table:bestfits} but using the fitting range $5<r<180~\hMpc$. Including smaller separations adds six data bins in the fits. The quality of the fits is marginally reduced compared to the fiducial fits.}
\label{table:bestfitsrmin5}
\end{table}

\bibliographystyle{JHEP}
\bibliography{qsocivDR14}

\end{document}